\documentstyle[preprint,tighten,floats,aps]{revtex}
\begin{document}
%
%
%		title
%
\preprint{IFUP-TH 21/93}
\draft
\title{
Scaling, asymptotic scaling and Symanzik improvement. Deconfinement
temperature in $SU\left(2\right)$ pure gauge theory.
}
\author{Giancarlo Cella\cite{auth::cella}, Giuseppe
Curci\cite{auth::curci},
Raffaele Tripiccione\cite{auth::lele},
Andrea Vicer\`e\cite{auth::vicere}}
\address{
Dipartimento di Fisica dell'Universit\'a di Pisa and \protect\\
Istituto Nazionale di Fisica Nucleare, I-56126 Pisa, Italy
}
\date{\today}
\maketitle
\begin{abstract}
We report on a high statistics simulation of $SU(2)$ pure gauge field
theory at finite temperature, using Symanzik action. We determine the
critical coupling for the deconfinement phase transition on lattices up to
$8 \times 24$, using Finite Size Scaling techniques.
We find that the pattern of asymptotic scaling violation is
essentially the same as the one observed with conventional, not improved
action. On the other hand, the use of effective couplings defined in terms
of plaquette expectation values shows a precocious scaling, with respect to
an analogous analysis of data obtained by the use of Wilson action, which we
interpret as an effect of improvement.
\end{abstract}
%
%
%	introduction
%
\section{Introduction}

In the last few years the increased computer power available for lattice
theorists has allowed a number of studies on relatively large lattices and
it has been possible to test accurately scaling and asymptotic scaling.

We have in mind in particular a series of works on $SU\left(2\right)$
lattice gauge theory at finite temperature
(\cite{fingberg:heller:karsch,engels:fingberg:miller} and references
therein) which have been able to test for asymptotic scaling on lattices up
to $N_\tau = 16$, showing still considerable scaling violations.

On the other hand, in the same simulations one observes a rather
good scaling of ratios of dimensionful quantities, thus supporting the idea
that asymptotic scaling is violated by universal terms, which cancel in the
ratio, and that some $\beta$ function, other than the perturbative one,
exists.

Since the Renormalization Group, parameterized in terms of the ``bare''
coupling on the lattice, shows large deviations from the asymptotic
behavior, we feel that it is very important to test different
renormalization schemes, as stressed in a recent work
by Lepage et al.~\cite{lepage:mackenzie}.
For instance, it has become more and more apparent that effective
schemes on one hand improve the asymptotic scaling~\cite{parisi1,parisi2},
and on the other hand reconcile the perturbative expansion of ultraviolet
dominated quantities with the lattice numerical
results~\cite{lepage:mackenzie}.

In this context, we have decided to test the effect of the use of a
different lattice formulation, using an improved Symanzik action in the
simulation of 4 dimensional lattice gauge theory.
We did not expect a dramatic improvement of asymptotic scaling, even if in a
pioneering work~\cite{curci:tripiccio} some indications of such an
improvement were found, but we have considered that by comparison with
similar works using Wilson action something about the origin of scaling
violations could be understood.

The study of this particular model, $SU\left(2\right)$ gauge theory at finite
temperature, is justified by its similarity with QCD, as well as by its
greater simplicity. Moreover, the particular observable chosen, the
temperature of deconfining phase transition, is well defined and can be
exactly determined by use of Finite Size Scaling (FSS) techniques, thus
allowing us to pursue our main objective, the study of scaling properties.

We can anticipate our results: the pattern of asymptotic
scaling violation observed with Symanzik action is essentially the same as
the one observed with the Wilson formulation, the only apparent effect is a
precocious transition from the strong to the weak coupling regime.
This result seems to indicate that a universal lattice $\beta$ function
exists.

Analogously the effective coupling approach, formulated in terms of plaquettes
expectation values, works well in improving the asymptotic scaling figure
and gives results consistent for the two lattice formulations.

In section\ \ref{symanzik} we give a short review of the Symanzik approach
to the lattice formulation of field theories and we recall physical
characteristics of the model under study.

In section\ \ref{fss}, we review the FSS techniques used in this work, in
order to give a presentation as self contained as possible.

In section\ \ref{simulation} we give details about our simulations.

In section\ \ref{results} we present the results of our measurements,
exploiting the FSS techniques to present various tests of consistency of
the results.

We discuss the results in section\ \ref{discussion}, together with an
analysis of the different renormalization schemes.

Appendix A is devoted to technical details on the subtraction of biases and
on error estimation.

Analogously in Appendix B we discuss our implementation of the Density of
States Method (DSM).
%
%
%	symanzik
%
\section{The Symanzik improvement and $SU\left(2\right)$ lattice gauge theory}
\label{symanzik}

\subsubsection{Generalities}

The formulation of a field theory on a lattice is a necessary step if we want
to study it in a non perturbative way with a computer simulation.
The lattice can be seen as an ultraviolet regulator as any other,
introducing in a natural way a short distance cut-off.

We can extract from a computer simulation relations between observables
(for example, mass ratios) only for this regularized theory, and
obviously we cannot recover completely the continuum limit, so our
predictions will be affected by systematic errors (the so-called
lattice artifacts).

The cut-off dependence of a theory can be reduced by a clever choice
of the regularization scheme: this is a well known technical point
which was studied extensively by
K.~Symanzik~\cite{symanzik:phi4,symanzik:sigma}. He realized that this fact
can be used to minimize the consequences of a non-zero lattice spacing $a$
(for an introductory discussion see the pedagogical work of
Parisi~\cite{parisi:sym}).

The key observation is that a field theory can be transcribed on a
lattice with considerable arbitrariness: the lattice action must only reduce
to the continuum one as $a \leadsto 0$. So we can redefine the action by
adding an arbitrary combination of irrelevant operators, which vanish in
this limit, and this is equivalent to change the regularization scheme.
Symanzik has also shown (see for instance~\cite{symanzik:eff}) that every
lattice regulated theory is perturbatively  equivalent to a local effective
lagrangian, in a given renormalization scheme, of the type
\begin{equation}
{\cal L}_{\text{latt}} \equiv {\cal L}_{\text{eff}} =
\sum_i c_i^{(0)} {\cal O}_i^{(0)} + a^2 \sum_i c_i^{(2)} {\cal O}_i^{(2)} +
a^4 \sum_i  c_i^{(4)} {\cal O}_i^{(4)} + \cdots,
\end{equation}
where ${\cal O}_i^{(n)}$ are local operators of dimension $d+n$, if $d$
is the dimension of space.
It follows that it is possible to build a lattice action in such a way that
the corresponding effective lagrangian is
\begin{equation}
\label{eq:limproved}
{\cal L}_{\text{latt,impr}} \equiv {\cal L}_{\text{eff,impr}} = {\cal
L}_{\text{cont}} + a^{2 p} \sum_i c_i^{(2 p)} {\cal O}_i^{(2 p)}  + \cdots,
\end{equation}
using an appropriate linear combination of the operators ${\cal O}_i^{(q<2p)}$
transcribed in an arbitrary way on the lattice.
The determination of the coefficients for this combination is a non trivial
task for an interacting theory: in principle it could be done with an high
precision numerical simulation.

We can also compute them perturbatively, if the bare coupling constant is
sufficiently small (i.e. near the continuum limit for an asymptotically free
theory), using a matching procedure between vertex
functions~\cite{symanzik:phi4}, or in the case of gauge theories (for
instance) matching the expectation value of gauge invariant quantities, like
the Wilson loops~\cite{weisz} or the inter-quark
potential~\cite{curci:menotti:paffuti}. A exposition of the different
improvement strategies can be found in the work of L\"uscher et
al.~\cite{luscher}.

If this program works, we may hope to obtain more accurate predictions for
physical quantities with the not extremely big lattices we can use with the
current computing resources.
The more direct way to verify that lattice artifacts are reduced is to measure
an adimensional ratio of physical quantities while making $a$ smaller and
smaller.
For example using the action in Eq.~(\ref{eq:limproved}) we expect to find
for a mass
ratio ($\xi$ = correlation length)
\begin{equation}
\label{eq:mratio}
\frac{m_a}{m_b} = K  \left( 1 + O\left[ \left( \frac{a}{\xi}\right)^{2 p + 2}
\ln\left(\frac{a}{\xi}\right)\right] \right)\ .
\end{equation}

A more refined test consists in verifying how well the quantities ${\cal M}_i$
we can measure on lattice obey the Renormalization Group equation, which
is merely a statement of cut-off independence of physical observables
(scaling). The lattice artifacts modify this equation adding non-universal
(i.e. ${\cal M}$-dependent) scaling-violating terms, so we have
\begin{equation}
\label{eq:rg}
\left( -a \frac{\partial}{\partial a} +
\left[ \bar{\beta}(g) + \bar{\beta}_{\cal M}(g) \right]
\frac{\partial}{\partial g}   \right)
{\cal M}_i = \Lambda_{\cal M}.
\end{equation}
Here $\bar{\beta}_{\cal M}$ is a non analytic contribution that cannot
be calculated perturbatively (for instance, in the bidimensional $\sigma$
model it can be evaluated in the large $N$ limit, where one finds
$\bar{\beta}_{\cal M} = O\left[g e^{-\frac{k}{g}}\right]$) and
$\Lambda_{\cal M} = O\left[\left(a/\xi\right)^2 \ln
\left(a/\xi\right)\right]$.
If we use the action (\ref{eq:limproved}) with $p=2$, we expect first of all
$\Lambda_{\cal M} = O\left[\left(a/\xi\right)^4 \ln
\left(a/\xi\right)\right]$, and then a reduced $\bar{\beta}_{\cal M}$.
In the real case we can calculate the improved action at best to the first
perturbative
order, and we expect contributions $O\left[g^4 \left(a/\xi\right)^2 \ln
\left(a/\xi\right)\right]$ to $\Lambda_{\cal M}$,
which give a little contribution for an asymptotically free theory if we
are sufficiently near the continuum.

\subsubsection{Scaling vs. asymptotic scaling.}

In the general case we know only the first few perturbative terms of the
function $\bar{\beta}(g)$, in particular the first two (scheme
independent) terms. These are the relevant terms in the continuum limit if
Asymptotic Freedom holds,
while in an intermediate coupling  range higher order contributions can be
important. If an observable ${\cal M}$ follows the renormalization group
with the perturbative approximation  to $\bar{\beta}(g)$ we say we are in
an asymptotic scaling region: if this is not true, it does not  imply that
the data cannot be trusted, but only that we had to use a better
approximation to the exact beta function. We can try to extract this
``improved'' $\bar{\beta}(g)$ from the simulation itself and then verify
the self consistency of this procedure. If we can find a function
$\bar{\beta}_{\text{eff}}(g)$ so that the RG evolution holds for different
quantities we say we are in the scaling regime.

It is important to note that
\begin{itemize}
\item{The Symanzik program can improve the scaling, but not necessarily the
asymptotic scaling: indeed, the onset of asymptotic scaling is expected in a
region where irrelevant operators give negligible contribution.}
\item{The scaling test is more conclusive than the stabilization of mass
ratios, as we can imagine that
there are some ${\cal M}$ independent contributions to $\Lambda_{\cal M}$
which cancel in Eq.~(\ref{eq:mratio}).}
\end{itemize}

As it is well known asymptotic scaling in the $SU\left(2\right)$ gauge
theory with the Wilson action has not yet been found (see for
instance~\cite{fingberg:heller:karsch}). So we must extrapolate the beta
function in some physically sensible way in the intermediate coupling
region.

\subsubsection{The model under study}

The $SU\left(2\right)$ gauge theory at finite temperature has been
discussed at length for instance in~\cite{engels:karsch:satz}: we here just
recall that the model shares with QCD the presence of a deconfining phase
transition, characterized by the breaking of the center $Z\left(2\right)$
symmetry~\cite{polyakov} and by the appearance of a non zero expectation
value for the thermal Wilson loop, or Polyakov line, $P$.
It is by now well established that the transition is of the second order
and all numerical simulations give results in good agreement with the early
ansatz~\cite{svetitsky:yaffe} that the model belongs to the same
universality class as the $3D$ Ising model.

In our simulations we have used a ``tree improved'' action, i.e. we
have only corrected in part lattice artifacts in the classical theory.

This can be done by observing that the usual Wilson action is equivalent to
the
effective lagrangian
\begin{equation}
{\cal L}_{\text{eff}} = -\frac{1}{4} F_{\mu \nu} F_{\mu \nu} +
\frac{1}{12} a^2 \partial_\mu F_{\mu \nu} \partial_\mu F_{\mu \nu} + O(a^4).
\end{equation}
We can compensate for the $O(a^2)$ term by adding a suitable irrelevant
operator. A  possible and widely used choice is
\begin{equation}
S_I = S_W + S_{\text{irr}} = \beta \sum U_{1\times1} +
\beta \sum \left( \frac{2}{3} U_{1 \times 1} - \frac{1}{12} U_{1 \times 2}
\right) =
\beta \sum \left( \frac{5}{3} U_{1 \times 1} - \frac{1}{12} U_{1 \times 2}
\right),
\end{equation}
where $U_{n \times m}$ is the $n \times m$ plaquette.

This choice has already been experimented in other works, both in the study
of the finite temperature theory~\cite{curci:tripiccio} and of the
quark-antiquark potential~\cite{gutbrod:montvay}, using the ichosaedral
approximation to the $SU\left(2\right)$ group. To our knowledge the
present work is the first where a 4-dimensional gauge theory is simulated
with the Symanzik action and the full group.
%
%
%	fss
%
\section{Finite Size Scaling analysis}
\label{fss}

The Finite Size Scaling (FSS) technique is by now a widely used tool in the
investigation of pseudo-critical properties of statistical systems in
finite volume. Its application to the analysis of second order
phase transitions permitted a detailed numerical test of theoretical
predictions on critical exponents. The effectiveness in the study of
$SU(2)$ deconfining phase transition has been already demonstrated in a
number of papers, where use of the Wilson's action was made
\cite{fingberg:heller:karsch,engels:fingberg:miller,%
engels:fingberg:mitrjushkin}.

In this section we collect some formulas needed by our investigation, in order
to give a presentation as self contained as possible of our results.
A more detailed general introduction can be found for instance in
~\cite{barber}, while we refer
to the original works~\cite{fingberg:heller:karsch,engels:fingberg:miller}
for the improvements of the method which we have applied to our analysis.

Numerical simulations of statistical systems are limited to finite volumes,
characterized by some length scale $L$. In the vicinity of an (infinite
volume) phase transition the system exhibits a ``pseudo-critical''
behavior, for instance at a second order phase transition the
susceptibility $\chi$ shows a peak broadened by the finite volume, while the
correlation length $\xi$, defined in terms of appropriate pair correlation
functions, reaches the dimensions of the system. For our purposes, the
shift in the critical temperature is the most important effect, which can
be easily estimated if one assumes some universality class for the model
under consideration and that the correlation length exhibits a power
behavior in the vicinity of the pseudo-transition.
\begin{equation}
\xi \propto \left|t\right|^{- \nu}
\end{equation}
where $t = \frac{T-T_c}{T_c}$ is the reduced temperature ( in units of the
infinite volume critical temperature ). If the finite volume ``critical''
temperature, $T_c\left(L\right)$, is reached when the correlation length is
of the order of the physical size of the system, it follows that the
expected shift is
\begin{equation}
\left|T_c\left(L\right) - T_c\right| \propto T_c L^{-\frac{1}{\nu}}\ .
\end{equation}
Note that in fitting measurements the critical temperature is a very
sensitive parameter, hence the error in its determination should be reduced
as much as possible if one is interested in determining universal
parameters with a good accuracy.

To extract infinite volume limits it is necessary an
``ansatz'' on the form of the observables.

The main idea is that near a second-order phase transition, as the
correlation length $\xi$ approaches the size $L$ of the system,  the
observables $O$ depend on this two scales only through the ratio $r =
L/\xi$, apart from a volume dependent prefactor
\begin{equation}
O_L = L^\omega \bar{Q}_0\left(r\right)
\end{equation}
or, in terms of the reduced temperature,
\begin{equation}
O_L = L^\omega \bar{Q}_0\left(L t^\nu\right)\ .
\end{equation}

If for large $L$ this observable exhibits a critical behavior, with
exponent $-\rho$, the function $\bar{Q}_O$ should behave like some power of
its argument, and the requirement of cancellation of volume dependence
requires that $\omega = \rho/\nu$. Hence it results the following
parameterization
\begin{equation}
O_L = L^{\rho/\nu} \bar{Q}_L\left(L t^\nu\right)\ ,
\end{equation}
or alternatively, and more commonly
\begin{equation}
O_L = L^{\rho/\nu} Q_L\left(t L^{1/\nu} \right)\ .
\end{equation}
The important point is that renormalization group arguments show
that all the critical behavior can be derived from the singular part of
the free energy density, which is then assumed to have the conventional
form
\begin{equation}
f_s = L^{-d} Q\left(g_T L^{1/\nu}, g_h L^{\left(\eta +
\gamma\right)/\nu}, g_i L^y_i\right)
\end{equation}
where $g_T,\,g_h$ are connected to the reduced temperature and the external
magnetic field by linear relations, plus corrections
\begin{eqnarray}
g_t &=& c_t t + O\left(t h, t^2\right)\nonumber\\
g_h &=& c_h h  + O\left(t h, h^2\right)\ .
\label{fss:scaling_fields}
\end{eqnarray}
The additional dependence on irrelevant scaling fields $g_i$, with
negative exponents $y_i$, cannot be neglected in many practical simulations,
and determines corrections to the scaling behavior. Note also that we name
$\eta$ the critical exponent of the ``magnetization'', instead of the
conventional $\beta$, to avoid confusion with the coupling in the
latticized theory.

Let us come to our system, and recall that we simulate on a lattice of
volume $L^3$, with $L = N_\sigma a$, and temporal extent $N_t$, determining a
temperature $T = 1 / \left(N_t a\right)$.

The relation between the spacing $a$ and the bare coupling $\beta = \frac{2
C_a}{g^2_0}$ is unknown, and one of the motivations of our simulation is the
study of the validity of the universal two-loop asymptotic formula
\begin{equation}
a \Lambda = \left(\frac{\beta}{2 C_a b_0}\right)^{b_1 / 2
b_0^2}\exp\left(-\frac{\beta}{4 C_a b_0}\right)\ .
\label{fss:asymptotic}
\end{equation}
As first noted in~\cite{fingberg:heller:karsch}, it is convenient to
rewrite the free energy density in terms of the dimensionless combination
$L T$, that is, of the ratio $y = N_\sigma / N_\tau$.
\begin{equation}
f_s\left(t,h;N_\sigma; N_\tau\right) =
y^{-d}
Q_f\left(g_ty^{\frac{1}{\nu}}, g_h
y^{\frac{\eta + \gamma}{\nu}}, g_i
y^{y_i}\right)\ .
\label{fss:f_sing}
\end{equation}
Define $P$ the spatial average of the Polyakov line
\begin{eqnarray}
L\left(\vec{x}\right) &=& \prod_{t=1}^{N_t} U\left(\vec{x},
t\right)\nonumber\\
P &=& N_\sigma^{-d}\sum_{\vec{x}} {\rm Tr}\left[L\left(\vec{x}\right)\right]
\end{eqnarray}
and introduce a source $h$ for this quantity in the partition function,
through the operator $h Z\left(a, N_\tau\right) N_\sigma^d P$, where the
renormalization factor $Z$ cancels the divergent self-energy contributions
to the Polyakov loop.

Derivatives with respect to the source $h$ allow to define the
physical order parameter\footnote{In a finite volume system, as tunnelling
between inequivalent ``vacua'' tends to restore the $Z\left(2\right)$
symmetry, the experimental order parameter is defined as $\left|P\right|$}
and the susceptibilities
\begin{eqnarray}
\left<P_p\right> &=& - \left.\frac{\partial f_s}{\partial h}\right|_{h=0} =
y^{-\eta/\nu}
Q_P\left(g_ty^{\frac{1}{\nu}}, g_i
y^{y_i}\right)\nonumber\\
\chi &=& \frac{\partial^2 f_s}{\partial h^2} = y^{\gamma/\nu}
Q_\chi\left(g_ty^{\frac{1}{\nu}}, g_i
y^{y_i}\right)
\label{fss:der_f}
\end{eqnarray}
where use has been made of the first hyperscaling relation
\begin{equation}
\gamma/\nu + 2 \eta/\nu = d
\label{fss:hyper1}
\end{equation}
implied by Eq.~(\ref{fss:der_f}).
\footnote{
In fact, assuming that fluctuations contribute to the singular part of the
free energy density as the ratio of the unit volume to the volume
$\xi\left(t\right)^d$
\[
f_s\left(t\right) \sim \xi\left(t\right)^{-d} \sim t^{\nu d}
\]
which is incorporated in Eq.~(\ref{fss:f_sing}), if for small values of $x$
\[
Q_f\left(x, 0, \dots\right) \sim \left(x\right)^{\nu d}\ .
\]
Now derivatives in Eq.~(\ref{fss:der_f}) can be written as
\begin{eqnarray*}
\left<P\right> &=& - \left(N_\sigma/N_\tau\right)^{-d + \left(\eta +
\gamma\right)/\nu} Q_1\left(g_t
\left(N_\sigma/N_\tau\right)^{1/\nu}\right)\nonumber\\
\chi &=& \left(N_\sigma/N_\tau\right)^{-d + 2 \left(\eta +
\gamma\right)/\nu} Q_2\left(g_t
\left(N_\sigma/N_\tau\right)^{1/\nu}\right)
\end{eqnarray*}
and so, combining the dependence of $Q_{1,2}$ on $g_t$ (that is, the same
as $Q_f$ ), with the requirement that in the infinite volume $N_\sigma
\rightarrow \infty$ limit the critical behavior is specified by the
exponents $\eta, \; \gamma$
\[
P \sim \left|t\right|^\eta\qquad\qquad \chi \sim \left|t\right|^{-\gamma}\; ,
\]
it results Eq.~(\ref{fss:hyper1})\ .
}

A useful quantity in the determination of the critical temperature
is then the Binder cumulant $g_4$, defined as the fourth derivative
of the free energy with respect to ``magnetic field'', normalized to the
susceptibility
\begin{equation}
g_4 = \left.\frac{\partial^4 f_s}{\partial h^4}\right|_{h=0}\left/
\left(\chi^2
\left(N_\sigma/N_\tau\right)^d\right)\right.\ .
\end{equation}
In fact, its expression is a directly scaling function, where explicit
dependence on $N_\sigma$ is canceled by the use of the hyperscaling relation:
moreover, the multiplicative renormalization $Z$ of the Polyakov loop
cancels, and therefore the expression for $g_4$ can be safely given in terms
of
expectation values of moments of Polyakov loop
\begin{equation}
g_4 = \frac{\left<P\right>^4}{\left<P\right>^2} - 3.0\ .
\label{fss:g4P}
\end{equation}
Its directly scaling form can be expressed as
\begin{equation}
g_4\left(t; N_\sigma; N_\tau\right) = Q_g\left(g_t\left(t,N_\tau\right)
y^{1/\nu},\; g_1 y^{y_1}\right)
\label{fss:g_4}
\end{equation}
where we have conserved the first irrelevant scaling field.
The localization of the critical point is then possible by finding the
intersections of $g_4$ curves, as functions of the temperature $t$, at
various values of the spatial size $N_\sigma$.

Taking into account the first irrelevant field one expands the expression
given in Eq.~(\ref{fss:g_4}) in the vicinity of the transition
\begin{equation}
g_4\left(t; N_\sigma; N_\tau\right) \simeq g_{4, 0} + g_{4, 1} t
y^\frac{1}{\nu} + g_{4, 2} y^{y_1}\ .
\end{equation}
Hence the intersection of two $g_4$ curves at values of the spatial size
$N_\sigma, N^\prime_\sigma$ corresponds to a shifted
temperature~\cite{engels:fingberg:miller}
\begin{equation}
t  = \frac{g_{4, 2}}{g_{4,1}}
\frac{\left[\left(y^\prime\right)^{y_1} -
y^{y_1}\right]}{\left[y^\frac{1}{\nu} -
\left(y^\prime\right)^\frac{1}{\nu}\right]} \propto
\left(\frac{N_\sigma}{N_\tau}\right)^{y_1 - 1/\nu} \frac{1 -
b^{y_1}}{b^{1/\nu} - 1}
\label{fss:t_shift}
\end{equation}
where $b = N_\sigma^\prime / N_\sigma$.

As $y_1 < 0$, this shift in the crossing goes to zero as $N_\sigma$ grows,
while also at fixed $N_\sigma$ as $b$ grows the shift goes down, governed
by the exponent $ - 1 / \nu$.

Actually, in the simulation of $SU(N)$ gauge theories, the connection
between the bare coupling $\beta$ and the lattice spacing, which allows to
determine the temperature in units of the reference scale $\Lambda$, is
known only in the asymptotic regime $g_0 \rightarrow 0$. As noted
in~\cite{fingberg:heller:karsch}, this means that
expressing the reduced temperature $t$ in terms of $\beta$ through an
approximate formula like
\begin{equation}
t = \left(\beta - \beta_{c,\infty}\right) \frac{1}{4 C_a b_0}\left[1 -
\frac{2 C_a b_1}{b_0} \beta_{c,\infty}^{-1}\right]
\end{equation}
introduces an error $O\left(\beta_{c,\infty}^{-1}\right)$ in the
determination of $t$.
This shift is of the order of $8\%$ in the relevant coupling regime
explored, but one should be aware that this estimate would be {\em a
posteriori} justified by the observation of an asymptotic scaling
behavior, which is absent.

In our search for the determination of the critical temperature this shift
is immaterial, as long as a relation of the form
\begin{equation}
t\propto \left(\beta - \beta_{c,\infty}\right)
\end{equation}
is valid.

Therefore Eq.~(\ref{fss:t_shift}) translates in the form
\begin{equation}
\beta_c\left(N_\sigma, N_\sigma^\prime\right) = \beta_{c,\infty} \left(1 -
c \varepsilon\right)
\label{fss:beta_shift}
\end{equation}
where
\begin{equation}
\varepsilon = N_\sigma^{y_1 - 1/\nu} \frac{1 - b^{y_1}}{b^{1/\nu} -
1}\qquad\qquad b = \frac{N_\sigma^\prime}{N_\sigma}
\end{equation}
and the constant $c$ maintains a dependence on $N_\tau$ and
$\beta_{c,\infty}$ which is irrelevant in the $\varepsilon \rightarrow 0$
extrapolation.
%

%
%	simulation
%
\section{Details of the simulation}
\label{simulation}

Most of the work has been done on the APE supercomputer~\cite{ape:machine}.
The model operating at Pisa is the so called ``tube'' machine,
a 128 processor parallel computer with a peak performance of 6 GigaFlops.

All the code has been written in the high-level
language~\cite{ape:language} proper of this machine (``apese''), and the
program runs at about $35\%$ of the peak speed: this result is quite good
if one takes into account the complications both in addressing and in
memory access due to the use of the Symanzik action. In other words, in our
code the floating point performance is slowed by the large number of
integer and addressing operations required to load the elements of group in
the register-file: this means that an analogous implementation for the
$SU\left(3\right)$ gauge group should result in a better performance, as
the ratio of floating point to integer operations would be increased both
in computation and in I/O.
A detailed account of the implementation will be given elsewhere.

In the update of lattices with $N_\tau = 3,4,\dots 8$ we have used an
overrelaxed Heat Bath update. An exact Heat Bath algorithm has been
implemented, in the modified Kennedy-Pendleton
form~\cite{kennedy:pendleton} which results in higher acceptance and is
well suited to a parallel machine like APE; a number of complete
overrelaxation sweeps ranging from $10$ to $16$ has been used to
decorrelate between subsequent Heat Bath sweeps.

As the Symanzik action couples next-to-nearest neighbors sites, on a
lattice $2 \times N_\sigma$ it results a self coupling of spatial links,
and therefore the action is quadratic and prevents the use of an Heat Bath
algorithm: hence on these smaller lattices we have used a Metropolis
algorithm running on RISC workstations.

In Tabb.~\ref{sim:rp1},\ref{sim:rp2},\dots \ref{sim:rp6} we present the
simulation parameters, together with the estimated autocorrelation time of the
Polyakov line. The Density of States Method has been used to interpolate
between simulated data points. See Appendix B for a discussion of the
meaning of the last column, where the ``reweighting range''
used in the interpolation is listed.

We have performed also a series of runs on large symmetric lattices to
measure the expectation values of plaquettes, $U_{1 \times 1},\,U_{1 \times
2}$, to be used in determining effective couplings ``a la Parisi''. We list in
Tab.~\ref{sim:rp7} the corresponding run parameters.
%
%
%	results
%
\section{Results}
\label{results}

\subsection{Determination of critical point by FSS techniques}

Runs on lattices with $N_\tau = 2,3,\dots 8$ and up to $N_\sigma = 24$ have
been devoted to explore the transition region and measure accurately
the Polyakov line.

\subsubsection{Binder method}

\begin{figure}
\protect{
\vskip 10truecm
\includegraphics{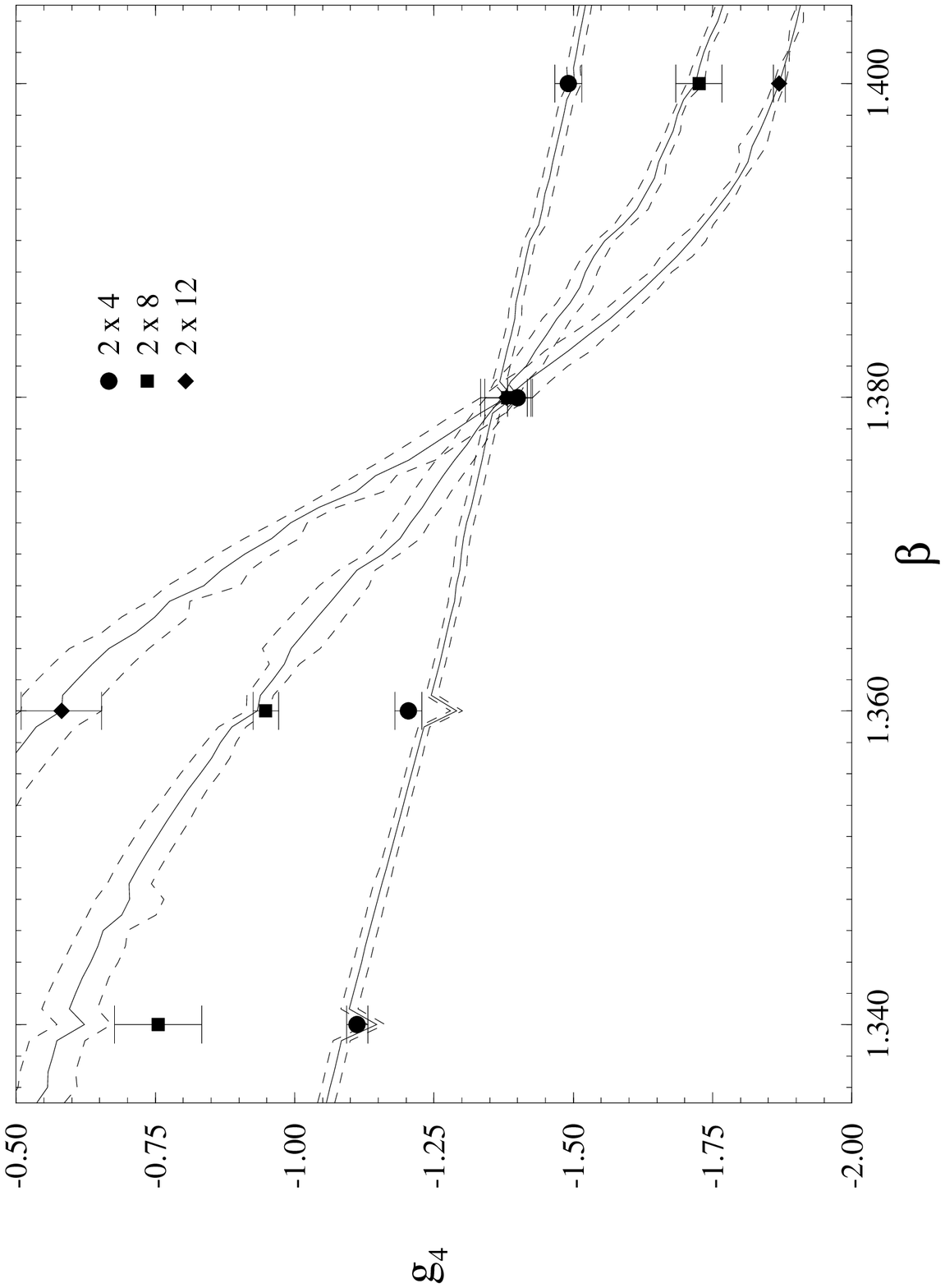}
}
\vskip 1.in
\caption{Binder cumulant: lattices with $N_\tau = 2$}
\label{fig1}
\end{figure}

\begin{figure}
\protect{
\vskip 10truecm
\includegraphics{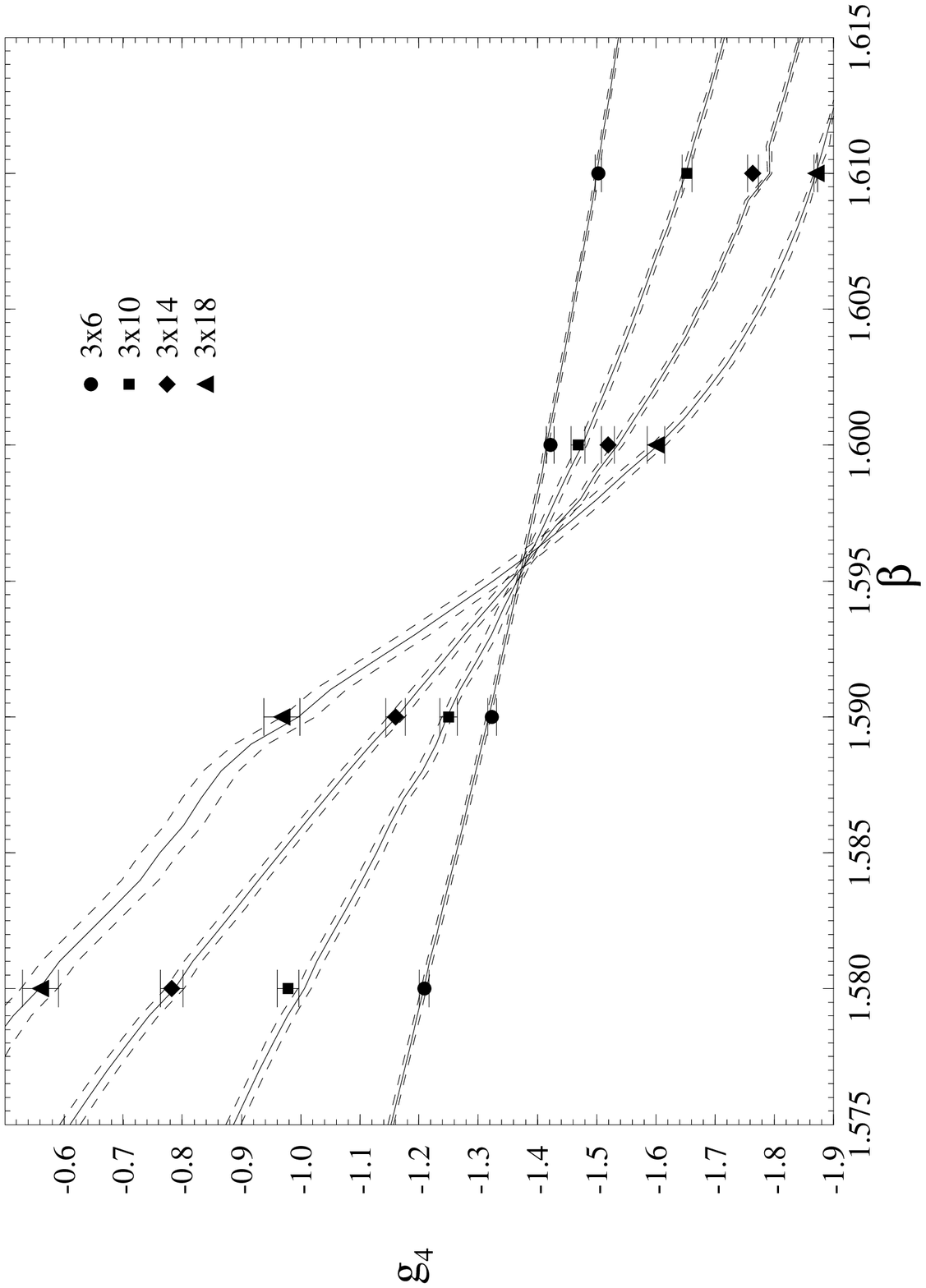}
}
\vskip 1.in
\caption{Binder cumulant: lattices with $N_\tau = 3$}
\label{fig2}
\end{figure}

In Figg.~\ref{fig1},\ref{fig2},$\dots$ \ref{fig7} we show the
plots of the Binder cumulant, Eq.~(\ref{fss:g4P}) for different values of
$N_\tau$ and $N_\sigma$.

As the Binder cumulant is not a self-averaging quantity, the resulting bias
has been computed, and found to be relevant for the larger lattices, where
statistic is relatively poor: details on the evaluation and subtraction of
the bias are reported in Appendix A.

The solid lines, as well as the dashed lines which are an estimate
of the error, are obtained with our implementation of the Density of
States Method~\cite{ferrenberg:swendsen:1,ferrenberg:swendsen:2}, described
in Appendix B.

The determination of the crossing points as well is based on DSM and we
report in Tab.~\ref{fss:crossings} the values of
intersections $\beta_c\left(N_\sigma, N_{\sigma^\prime}\right)$ for the
different lattices.

We make use of the extrapolation in Eq.~(\ref{fss:beta_shift}) based on the
introduction of a single irrelevant field, with exponent $y = -1$, to obtain
for lattices $3-6$ the infinite volume limit of the critical coupling:
\begin{equation}
\begin{array}{lll}
N_\tau = 3 \hskip 1truecm& \beta_{c,\infty} = 1.59624(13) &\hskip 1truecm
a = 0.23(5)\\
N_\tau = 4 \hskip 1truecm& \beta_{c,\infty} = 1.699(1)    &\hskip 1truecm
a = 0.20(6)\\
N_\tau = 5 \hskip 1truecm& \beta_{c,\infty} = 1.76948(3)  &\hskip 1truecm
a = 0.85(10)\\
N_\tau = 6 \hskip 1truecm& \beta_{c,\infty} = 1.8287(11)  &\hskip 1truecm
a = 3(1)\ .
\end{array}
\end{equation}
For what concerns the lattice with $N_\tau = 2$, we have not enough
statistics to distinguish the different intersection points. As this point
is in the strong coupling region, an high precision determination is not
necessary.

For the $N_\tau = 7,8$ lattices we have only a single crossing, so we can
estimate the critical coupling by assuming some value for the $a$ coefficient.
By choosing $a \simeq 3-5$ for the $N_\tau = 7$ lattice we obtain, having
$b = 20/14 = 1.43$, a value for $\varepsilon \simeq 4.2 \times 10^{-4}$, hence
$a\varepsilon \simeq 1-2 \times 10^{-3}$. Analogously one obtains
$a\varepsilon
\simeq 3 \times 10^{-3}$ for the $N_\tau = 8$ lattice, so we estimate a
systematic error of order $10^{-3}$, giving so
\begin{equation}
\begin{array}{ll}
N_\tau = 7 \hskip 1truecm&  \beta_{c,\infty} = 1.8747 \pm 0.002 \pm 0.002\\
N_\tau = 8 \hskip 1truecm&  \beta_{c,\infty} = 1.920 \pm 0.004 \pm 0.003\ .
\end{array}
\end{equation}

Using the two loop asymptotic formula, Eq.~(\ref{fss:asymptotic}), this values
are reexpressed in terms of the $T / \Lambda_{\text{I}}$ ratio in
Tab.~\ref{fss:tc} .

\subsubsection{$\chi$ method}

A different way to determine the critical point has been recently proposed
by Engels et al. in \cite{engels:fingberg:mitrjushkin}.
The idea is to exploit the specific form of the scaling law for the
susceptibility, which can be written as follows
\begin{equation}
\chi_v = N_\sigma^3 \left<P^2\right>\ ,
\label{fss:true_chi}
\end{equation}
taking into account that in finite volume the expectation value of Polyakov
loops is set to zero by spin flips between degenerate vacua.

Expanding in the vicinity of the transition the expression in
Eq.~(\ref{fss:der_f}) one obtains
\begin{eqnarray}
\chi_v &=& \left(\frac{N_\sigma}{N_\tau}\right)^{\gamma/\nu}\left\{c_0 +
\left(c_1 + c_2
\left(\frac{N_\sigma}{N_\tau}\right)^{y_1}\right) t
\frac{N_\sigma}{N_\tau}^{1/\nu} + c_3
\left(\frac{N_\sigma}{N_\tau}\right)^{y_1}\right\}\nonumber\\
\left.\ln\left(\chi_v\right)\right|_{t = 0} &=& \ln\left({c_0}\right) +
\frac{\gamma}{\nu} \ln\left(\frac{N_\sigma}{N_\tau}\right) + \frac{c_3}{c_0}
\left(\frac{N_\sigma}{N_\tau}\right)^{y_1}\ .
\label{fss:true_chi_scaling}
\end{eqnarray}
Hence, assumed that the effect of the irrelevant is small ($y_1 \simeq
-1$), at the phase transition the true susceptibility $\chi_v$, plotted as
a function of the spatial size on a doubly logarithmic scale, should
exhibit a linear behavior.

Let us apply this method to the $N_\tau = 3$ lattice. We use the density of
states method to determine $\chi_v$ at various values of $\beta$, and
fit the data to a doubly logarithmic law.
The resulting value for the minimum of $\chi^2$ is at $\beta_c = 1.5956(4)$
at $95\%$ c.l., which is in good agreement with the $g_4$ determination.
One obtains also the slope $\gamma / \nu = 1.91(3)$, which exhibits a $3\%$
difference from the known value in the $3D$ Ising model $\gamma / \nu =
1.965(11)$.

The same procedure has been applied to the other lattices: we find that,
when applicable to our data, this method gives essentially the same result
as the method of Binder cumulant, so we have used it as a cross check of
the validity of the analysis.

\subsubsection{Universal Scaling Behavior}

We may test the universal scaling laws, expressed by Eqq.~(\ref{fss:g_4})
and (\ref{fss:der_f}), by plotting $g_4\left(t, y\right)$, ( $y = N_\sigma /
N_\tau$ ), as a function of the combination $t y^{1/\nu}$, and the analogous
expression for the Polyakov loop, $N_\tau
y^{\eta/\nu}\left<\left|P\right|\right>$: we choose $\nu = 0.628,\
\eta/\nu = 0.516,\ \gamma/\nu = 1.965$ as given by the $3D$ Ising model.
In Fig.~\ref{fig8} one may see that the $g_4$ values cluster
round a straight line, thus supporting the theoretical conclusions. The
agreement is not so good for the Polyakov line, Fig.~\ref{fig9},
which is sensitive to the self energy contribution.

\subsection{Critical temperature in the bare scheme}

Let us give in Fig.~\ref{fig10} a cumulative plot of the various
determinations of the critical temperature. The data on Wilson action are
taken from~\cite{fingberg:heller:karsch} and normalized to the point at
$N_\tau = 8$, ( instead of using, for instance, the perturbative $\Lambda$
ratio).

A few comments are in order:
\begin{itemize}
\item the weak coupling regime appears to be reached earlier in the
simulation with the Symanzik action,
\item in the weak coupling regime the pattern of asymptotic scaling
violation appears to be essentially the same, by using Wilson and Symanzik
action.
\end{itemize}

This result is in a clear disagreement with the earlier work by Curci et
al.~\cite{curci:tripiccio}, where asymptotic scaling was found with the
Symanzik action already at $N_\tau = 5$. We have then analyzed our data
with the same method, based on a fit to the critical behavior of the
Polyakov line. For $\beta > \beta_c$, the form
\begin{equation}
\left<\left|L\right|\right> = A\;\left(\beta - \beta_c\right)^{\eta}
\label{fitform}
\end{equation}
is assumed, while for $\beta < \beta_c$ the Polyakov line is set to zero:
this introduces some arbitrariness in the fitting procedure. In fact, using
Eq.~(\ref{fitform}) to determine $\beta_c$ one is forced to discard data
points assumed to be on the ``left'' of the transition. Moreover, on the
``right'' side of the transition (smaller lattice spacing) there are
stronger renormalization effects, and these introduce a systematic which
could be reduced by subtracting perturbative tails.

In Fig.~\ref{fig11} we show the data points, together with the results of a
three parameters fit to the critical behavior: the study is done for
lattices $3 \times 10, 4 \times 12, 5 \times 16$.

Results of the fit are shown in Tab.~\ref{fittab}, together with the value
of the best reduced $\chi^2$ found. Errors are estimated with the method of
$\chi^2$ equisurfaces.

The resulting critical temperatures are also reported in
Fig.~\ref{fig10}.
The high value of the reduced $\chi^2$, especially
on the $4 \times 12$ lattice, shows the bad quality of the fit, but for our
purposes it is important to point out that the finite volume ``critical''
temperature determined by this method shows no sign of asymptotic scaling.
On the contrary, it is present only a constant shift from the infinite
volume limit, and we conclude that the results of ~\cite{curci:tripiccio}
were hampered by a insufficient statistic.

\subsubsection{Perturbative $\Lambda$ ratio}

We can exploit the high precision data obtained with the Wilson action by
Fingberg et al.~\cite{fingberg:heller:karsch} to give in
Tab.~\ref{lambda_ratios} the resulting determination of $\Lambda$ ratios,
by using the asymptotic formula
\begin{equation}
\frac{\Lambda_{\text{I}}}{\Lambda_{\text{L}}} =
\left(\frac{\beta_{\text{I}}}{\beta_{\text{L}}}\right)^{b_1 /
2b_0^2} \exp\left(\frac{\beta_{\text{L}} - \beta_{\text{I}}}{4 C_a
b_0}\right)\ .
\end{equation}

These values are to be confronted with the perturbative results from Weisz
and Wohlert~\cite{weisz:wohlert}
\begin{equation}
\frac{\Lambda_{\text{I}}}{\Lambda_{\text{L}}} = 4.13089(1)\ .
\end{equation}
showing a discrepancy of the order of $10\%$.
%
%
%	discussion
%
\section{Discussion of results}
\label{discussion}

We can resume the outcome of our simulations as follows:
Symanzik and Wilson actions give consistent results in the so called
``bare'' scheme, that is, the Renormalization Group evolution driven by
scheme independent part of the perturbative $\beta$ function does not seem
to show scaling of the temperature. The pattern of scaling violation is
similar, at least for $N_\tau > 4$.

On the other hand it is well known that adimensional ratios of physical
quantities show a good scaling in the
coupling range considered in this work, at least in the case of Wilson
action, where the ratio $T_c/\sigma^{1/2}$ has been showed to scale over
the entire $\beta=2.30 - 2.74$ range~\cite{fingberg:heller:karsch}.

In our opinion these two facts, that is, the good scaling and the similar
pattern of asymptotic scaling violation in the two actions, are consistent:
as anticipated, in the scaling region the improvement has no effect on the
asymptotic scaling.

Let us first elaborate on the statement that the two scaling figures are
compatible with each other: to do this, we set up a perturbative framework and
discuss the effect of the coupling redefinition implied by the use of
Symanzik action.

We split $\beta(g)$ in the universal $\beta_{u} = - b_0 g^3 -
b_1 g^5$ and in the scheme dependent (unknown) part $\beta_{s.d.} = O(g^7)$,
where
\begin{equation}
b_0 = \frac{1}{\left(4\pi\right)^2}\frac{11 C_a}{3}\qquad\qquad
b_1 = \frac{1}{\left(4 \pi\right)^4}\frac{34 C_a^2}{3}\ .
\end{equation}
The beta function determines the dependence of lattice spacing $a$ on the $g$
value:
\begin{eqnarray}
\label{eq:solvebeta}
\ln\left(a\right) & = &  - \int \frac{dg}{\beta(g)} + C = \int \left(
\frac{1}{b_0 g^3} -
\frac{b_1}{b_0^2 g} + \sum_{n=0}^{\infty} c_n g^{2n + 1}  \right) dg +
C\nonumber\\ &=&  - \frac{1}{2 b_0 g^2} - \frac{b_1}{b_0^2} \ln(g) +
\sum_{n=0}^{\infty} c_n \frac{g^{2n + 2}}{2n + 2} , \nonumber \\
a(g) & = & \frac{1}{\Lambda_L} \left( b_0 g^2 \right)^{- \frac{b_1}{2 b_0^2}}
\exp \left( - \frac{1}{2 b_0 g^2} \right)
\left( 1 + g^2 \sum_{n=0}^{\infty} d_n g^{2 n}  \right)\ .
\end{eqnarray}
The coefficients $d_n$ in Eq.~(\ref{eq:solvebeta}) are generated by
$\beta_{s.d.}$; they play the role of asymptotic scaling violating terms.
For instance, in terms of the first sub-asymptotic term in $\beta_{s.d.}$,
the $b_2$ coefficient, one easily finds
\begin{equation}
d_0 = \frac{1}{2 b_0}\left(\frac{b_2}{b_0} -
\left(\frac{b_1}{b_0}\right)^2\right)
\end{equation}
If we redefine the coupling constant
\begin{equation}
g = g(\bar{g}) = \bar{g} + \alpha_1 \bar{g}^3 + \alpha_2 \bar{g}^5 + \cdots +
\alpha_n \bar{g}^{2 n + 1} + \cdots
\end{equation}
we obtain
\begin{equation}
a(\bar{g})  =  \frac{e^{\alpha_1/b_0}}{\Lambda_L}
\left( b_0 \bar{g}^2 \right)^{- \frac{b_1}{2 b_0^2}}
\exp \left( - \frac{1}{2 b_0 \bar{g}^2} \right)
\left( 1 + \bar{g}^2 \sum_{n=0}^{\infty} d_n^\star \bar{g}^{2 n}  \right)\ .
\end{equation}
This is the evolution dictated by a new beta function $\beta^\star(g)$,
\begin{equation}
\beta^\star(\bar{g}) = \beta(g(\bar{g}))
\left( \frac{\partial g(\bar{g})}{\partial \bar{g}} \right)^{-1} =
\beta_u(\bar{g}) + \beta^\star_{s.d.}(\bar{g})\ ,
\end{equation}
and the non universal terms are changed, for instance one has
\begin{equation}
b_2^\star =  b_2 - 2 \alpha_1 b_1 - 3 \alpha_1^2 b_0 + 2 b_0 \alpha_2\ .
\end{equation}
The prefactor $\exp\left(\alpha_1 / b_0\right)$ redefines the scale
parameter, consequently
\begin{equation}
\alpha_1 = - b_0 \ln\left(\frac{\Lambda^\star}{\Lambda_L}\right)\ .
\end{equation}
On the other hand the asymptotic scaling violating contribution is changed
\begin{equation}
d_0^\star =
d_0 - \alpha_1 \left(\frac{b_1}{b_0^2} + \frac{3\alpha_1}{2 b_0}\right) +
\frac{\alpha_2}{b_0}\ ,
\end{equation}
and one may hope to reduce its effect by a clever (and physically
motivated) choice of $\bar{g}$.

To study if the asymptotic scaling violation can be ascribed to the $d_2$
terms, we have performed an exploration of values of $b_2$ for both actions:
in Fig.~\ref{fig12} we show how setting values of $b_2$ for Wilson and
Symanzik actions respectively to $0.0018, 0.0011$, one is able to obtain a
good scaling: the $N_\tau = 2$ term has been dropped for clarity.

On the other hand, we are well aware of the arbitrariness of the procedure,
as we have checked that this fit procedure is unstable if also the coefficient
$b_3$ is introduced. This is easily understood as values of $b_2$ so large
are hard to justify on a perturbative ground: too see this consider the
running
coupling equation in terms of the characteristic QCD coupling, $\alpha_s /
\left(4\pi\right)$
\begin{eqnarray}
a\frac{d}{d\,a} \left(\frac{\alpha_s}{4\pi}\right) &=& \frac{22 C_a}{3}
\left(\frac{\alpha_s}{4\pi}\right)^2 + \frac{68 C_a^2}{3}
\left(\frac{\alpha_s}{4\pi}\right)^3 + \left(4\pi\right)^6 b_2
\left(\frac{\alpha_s}{4\pi}\right)^4 + \dots\nonumber\\
&\simeq& \frac{44}{6} \left(\frac{\alpha_s}{4\pi}\right)^2 + \frac{272}{3}
\left(\frac{\alpha_s}{4\pi}\right)^3 + 4330
\left(\frac{\alpha_s}{4\pi}\right)^4 +\dots\ :
\label{eq:alphaevolution}
\end{eqnarray}
for instance at $\beta \simeq 1.9$ one has $\alpha_s / \left(4\pi\right)
\simeq 0.013$, which means that the last term in Eq.~(\ref{eq:alphaevolution})
is of the same order of magnitude as the preceding one. So we can merely
regard this procedure as the definition of an effective $b_2$, much in the
spirit of the effective $\Lambda$ used in~\cite{ma:wetzel}.

\subsubsection{Perturbative approach: $\alpha_v$ coupling}

Let us consider more closely the poor convergence of the speculative
expansion given in Eq.~(\ref{eq:alphaevolution}): in the preceding
subsection we have looked for an appropriate $b_2$ coefficient without any
physical motivation.
On the other hand, Lepage and Mackenzie~\cite{lepage:mackenzie} point out
how a bad choice of the expansion parameter may give rise to a poorer
convergence of the perturbation series, and they suggest strategies to
design improved couplings, defined in terms of physical quantities.

This is much in the spirit of the old suggestions of Parisi, to define an
effective coupling by inverting the perturbative expression for some
ultraviolet-dominated quantity, like the single plaquette.

In~\cite{lepage:mackenzie} it is suggested to define an effective coupling
$\alpha_v$ in terms of the quark-quark potential, by the relation

\begin{equation}
V\left(q\right) = - \frac{C_f \left(4\pi\right)
\alpha_v\left(q^2\right)}{q^2}\ .
\end{equation}

This coupling can be perturbatively related to $\alpha_0$, the bare lattice
coupling, by the formula

\begin{equation}
\alpha_v\left(q^2\right) = \alpha_0\left(1 +
\alpha_0\left(\left(4\pi\right) b_0\ln\left(\frac{\pi}{a q}\right)^2 +
K\right)\right)\ ,
\end{equation}

where the constant $K$ can be extracted by Kovacs
work~\cite{kovacs}: she reports

\begin{equation}
V\left(q\right) \propto -
\frac{1}{q^2}\left(b_0\ln\left(\frac{q^2}{\pi^2 \Lambda_L^2}\right) +
\frac{C_a}{\left(4\pi\right)^2} J_L\right)^{-1}\ .
\end{equation}
Numerically one finds
\begin{equation}
J_L = \left\{\begin{array}{l}
- 16.954,\;\text{for}\; N = 2 \\
- 19.695,\; \text{for}\; N = 3\end{array}\right.\ .
\end{equation}

To translate the result for the Symanzik action it is sufficient to
rescale $\Lambda$ factors and impose equality in the physical result.
The defining relation is
\begin{equation}
b_0\ln\left(\frac{1}{\Lambda_L^2}\right) + \frac{C_a}{\left(4\pi\right)^2}
J_L \equiv
b_0\ln\left(\frac{1}{\Lambda_I^2}\right) + \frac{C_a}{\left(4\pi\right)^2}
J_I\ .
\end{equation}
One obtains easily, for $SU(2)$,
\begin{equation}
J_I = J_L + \frac{11}{3}\ln\left(\frac{\Lambda_I}{\Lambda_L}\right)^2\ .
\end{equation}
It is then possible to use the result of Weisz and
Wohlert~\cite{weisz:wohlert} on the $\Lambda$ ratio:
\begin{eqnarray}
\frac{\Lambda_I}{\Lambda_L} &=&
\exp\left[\frac{1}{\left(4\pi\right)b_0}\left(C_a\left(0.04329017(1)\right) -
\frac{1}{C_a}\left(0.04141417(1)\right)\right)\right]\nonumber\\
&=& \left\{\begin{array}{l}4.13089(1),\;\text{for}\; N = 2\\
5.29210(1),\;\text{for}\; N = 3\ ,
\end{array}\right.
\end{eqnarray}
and it follows
\begin{equation}
J_I = \left\{\begin{array}{l}-6.54859,\;\text{for}\; N = 2\\
-7.47609,\;\text{for}\; N = 3\ .\end{array}\right.
\end{equation}
Translating into the expression for $K$, one easily obtains
\begin{eqnarray}
K_L &=& \left\{\begin{array}{l}2.69831,\;\text{for}\; N =
2\\4.7018,\;\text{for}\; N = 3\end{array}\right.\nonumber\\
K_I &=& \left\{\begin{array}{l}1.04224,\;\text{for}\; N =
2\\1.78479,\;\text{for}\; N = 3\ .\end{array}\right.
\end{eqnarray}

We can now use value of $\alpha_0$ to set the value
$\alpha_v\left(\pi/a\right)$: one has obviously
\begin{equation}
\alpha_v\left(\pi/a\right) = \alpha_0\left(1 +
\alpha_0 K_{L\left(I\right)}\right)\ .
\end{equation}

For a given value $\beta_c = C_a/\left(2 \pi \alpha_c\right)$ of the
measured critical coupling, the corresponding critical temperature in the
$\alpha_v$ scheme may be obtained by using the relation
\begin{equation}
\frac{T}{\Lambda_v} = \frac{1}{\pi N_\tau}\frac{\pi}{a\Lambda_v}
\end{equation}
and the two loop scaling formula

\begin{equation}
\frac{q}{\Lambda_v} = \exp\left[\frac{1}{\left(8\pi\right)
b_0}\frac{1}{\alpha_v\left(q\right)} - \frac{b_1}{2
b_0^2}\ln\left(\frac{1}{4\pi
b_0\alpha_v\left(q\right)}\right)\right]\ .
\label{lepage:sf}
\end{equation}

Let us report in Tabb.~\ref{lepage:TcW} and \ref{lepage:TcS} the results of
the
analysis both for the Wilson and the Symanzik case.

It appears that the use of a perturbative expression for the effective
coupling $\alpha_v$ does not give a better asymptotic scaling. It should be
stressed however that this was merely an exercise, as the Lepage scheme
is intended to work in conjunction to a non perturbative determination of
the coupling $\alpha_v$, based on the measurement of the heavy quark
potential.
% As we do not have data on the heavy quark potential, obtained by the use of
% the Symanzik action, we now turn to the approach based on a different UV
% dominated quantity, the plaquette expactation value.

\subsubsection{Non-perturbative approach}

The bare coupling itself has no meaning in itself, so it must be eliminated
in favor of some physical quantity, connected to it.
In the perturbative approach this is a way to hide infinities and obtain a
renormalized field theory.

A simple way to implement this program on the lattice is the elimination of
the bare coupling in favor of some UV dominated quantity, such as the
expectation value of plaquettes. To this end, let us recall the results
of the work of Weisz and Wohlert~\cite{weisz:wohlert} on their perturbative
expansion: by defining, for simplicity in the adjoint representation
\begin{equation}
\ln\left[\frac{1}{C_A}\left<\text{Tr}\,U(L,T)\right>\right] =
-\sum_{n=1}^\infty \frac{g^{2n}}{\left(2 n\right)!} w_n\left(L, T\right)\ ,
\end{equation}
the first coefficient is given by
\begin{equation}
w_1\left(L,T\right) = C_f I\left(L, T\right)
\end{equation}
and the following values are given
\begin{eqnarray}
I\left(1, 1\right) &=& \left\{{\begin{array}{lr} \frac{1}{2} &
\text{Wilson} \\ 0.366262 & \text{Symanzik}\end{array}}\right.\nonumber\\
I\left(1, 2\right) &=& \left\{{\begin{array}{lr} 0.862251 & \text{Wilson}
\\ 0.662624 &\text{Symanzik}\end{array}}\right.\ .
\end{eqnarray}
Coefficients for the Wilson case are known up to fourth $w_4$, while for
Symanzik we shall limit ourselves to lowest order.

This allows to define two effective couplings in the Symanzik case,
depending on the two expectation values measured. That is, knowing that
\begin{eqnarray}
\ln\left[\frac{1}{C_A}\left<\text{Tr}\,U(L,T)\right>\right] = - \frac{g^2}{2}
C_f I\left(L, T\right)
\end{eqnarray}
and given $\beta = \frac{2 C_a}{g^2}$, we define
\begin{equation}
\beta_{1(2)} = C_a C_f \frac{I\left(1,1\left(2\right)\right)}{1 - U _{1
\times 1 \left(1 \times 2\right)}}\ .
\end{equation}
Numerically, for the $SU(2)$ case, this corresponds to
\begin{equation}
\beta_{1(2)} = \frac{0.549393 \left(0.993936\right)}{1 - U _{1
\times 1 \left(1 \times 2\right)}}\ .
\end{equation}
By using the results of measurements on large symmetric lattices we can
obtain via interpolation the values of plaquettes needed to compute the
effective coupling: we report the results in Tab.~(\ref{eff:data}).

We plot in Fig.~\ref{fig13} the comparison of the different schemes,
together with the Wilson data, taken from~\cite{fingberg:heller:karsch}, in
the effective scheme deduced from the single plaquette. As usual we
normalize to the last point: we drop the point at $N_\tau = 2$ to show in a
larger scale the other points.
It is worth noting that
\begin{itemize}
\item the two effective schemes used for the Symanzik action agree with
each other,
\item they both show a good scaling starting at $N_\tau = 4$,
\item the same scheme used with Wilson action seems to show, starting at
$N_\tau = 5$ the same behavior.
\end{itemize}

We think that these results can be interpreted as a confirmation of the
value of effective schemes, and as an indication that Symanzik action gives
an enlargement of the scaling window.
%
%
%	conclusions
%
\section{Conclusions}
\label{conclusions}

In this work we have performed an high statistics simulation of the $SU(2)$
pure gauge theory at finite temperature, using the tree-improved Symanzik
action and exploring a range of lattices up to $N_\tau = 8$; we have
determined, with the help of Finite Size Scaling methods,
the critical couplings for the deconfinement transition,
with the purpose of studying the scaling properties of the critical
temperature and the effect of improvement.

We can resume as follows our findings:
\begin{itemize}
\item asymptotic scaling violations are present, and are of the same size
as in the case of Wilson action. The pattern of violation is similar and
can be interpreted as driven by the same lattice $\beta$ function, modulo
scheme redefinitions.
This means either that lattice artifacts are small, or they are the same
for the two actions, in accordance with previous determinations of the
scaling window.
\item The use of a non perturbative coupling derived from the plaquette
expectation value gives a better scaling figure: this effective coupling
was already used in the same context but with the Wilson action, and
comparison of the two analyses shows that Symanzik action apparently gives
a slightly precocious scaling.
\item In our study we are unable to directly test the scaling, so it could
be desirable a zero temperature study of some other dimensionful quantity,
like masses or the string tension.
\end{itemize}

From a practical point of view, if we are willing to trust the precocious
onset of scaling in the effective scheme, we can say that the gain in
volume obtained by use of the improved action compensates the increased
complexity in the update routine, thus slightly favouring the use of this
technique in simulations. This should be even more true in the case of
$SU\left(3\right)$ group.

\acknowledgments

We thank Prof. Paolo Rossi for many useful discussions, and for a careful
reading of the manuscript.

We acknowledge the continuous support of many people from the Rome APE
group, particularly Simone Cabasino and Gian Marco Todesco.
%
%
%	references
%

%
%
%	appendixes
%
\appendix
\section{Statistical evaluation of Binder's cumulant.}
\label{binder}

In order to extract in a reliable way the quantities we are interested
in from a set of $N$ measures we have to correct for the distortions
induced by the finite sampling.
It is well known that often the ``naive'' estimator for a statistical
quantity is not the best one.
The much simpler and well known example is the estimator of variance
$\sigma^2$ for a set of $N$ uncorrelated data $x_i$. One may try
\begin{equation}
E_{b}(\sigma^2) = E( < x^2 > ) - E( <x> )^2 = \left( \frac{1}{N}
\sum_{i=1}^{N} x_i^2 \right)
- \left( \frac{1}{N} \sum_{i=1}^{N} x_i \right)
\left( \frac{1}{N} \sum_{j=1}^{N} x_j \right)
\end{equation}
but, owing to the fact that $E( <x> )$ is a random variable with a
finite variance the estimator's expectation value is
\begin{equation}
< E_{b}(\sigma^2) > = \frac{1}{N} \sum_{i=1}^{N} < x^2 > - \frac{1}{N^2}
\sum_{i=1,j=1}^{N} \left( \delta_{ij} <x^2> + (1-\delta_{ij}) <x>^2
\right) = \left( 1 - \frac{1}{N} \right) \sigma^2,
\end{equation}
which is ``biased'' by a finite sample effect $O(\frac{1}{N})$.
To evaluate in the best way $\sigma^2$ we must correct for this bias:
this can be done using a new redefined estimator
\begin{equation}
E(\sigma^2)~=~N/(N-1)~E_{b}(\sigma^2)
\end{equation}
which gives the correct result $< E(\sigma^2) > = \sigma^2$.

Now we want to study the Binder's cumulant $g_4$ of a random variable $L$,
where the general cumulant $g_{2n}$ (in the $<L^{2n+1}>=0$ case which is of
interest to us) is defined as
\begin{eqnarray}
g_{2n} & = & \frac{<L^{2n}>_c}{<L^2>_c^n}, \\
g_4 & = & \frac{<L^4>_c}{<L^2>_c^2} = \frac{<L^4>}{<L^2>^2} - 3, \\
g_6 &=& \frac{<L^6> - 15 <L^2> <L^4> + 30 <L^2>^3 }{<L^2>^3}
\end{eqnarray}

The ``natural'' estimator for $g_4$ over a set of $N$ measurements is given by
\begin{equation}
\label{eq:naiveestimator}
\bar{g}_4 = \left(\frac{1}{N} \sum_i L_i^4\right) \left(\frac{1}{N} \sum_i
L_i^2\right)^{-2} - 3,
\end{equation}
but as we will see this is not a self-averaging quantity, in the sense that
$<\bar{g}_4> \neq g_4$, and we have to redefine it.
To calculate the bias of $g_4$ we introduce the $O(\frac{1}{\sqrt{N}})$
fluctuations $\eta_2$ and $\eta_4$ defined by
\begin{eqnarray}
\eta_2 & = & \frac{1}{N} \sum_i L_i^2 - <L^2> \nonumber \\
\eta_4 & = & \frac{1}{N} \sum_i L_i^4 - <L^4>
\end{eqnarray}
and we expand Eq.~(\ref{eq:naiveestimator}) around $\eta = 0$, to obtain
\begin{eqnarray}
\label{eq:fluctuationsg4}
\bar{g}_4 & = & g_4 \left[ 1 - 2 \frac{\eta_2}{<L^2>} + \frac{\eta_4}{<L^4>} +
3 \frac{\eta_2^2}{<L^2>^2} - 2 \frac{\eta_2 \eta_4}{<L^2><L^4>} +
o\left( \frac{1}{N} \right) \right].
\end{eqnarray}
Now to evaluate the corrections we take the expectation value, and we
get
\begin{equation}
\label{eq:g4meanbiased}
<\bar{g}_4> = g_4 \left[ 1 + \frac{3}{<L^2>^2} <\eta_2^2> -
\frac{2}{<L^2><L^4>} <\eta_2 \eta_4> + o\left(\frac{1}{N}\right) \right]
\end{equation}
where
\begin{eqnarray}
<\eta_2 \eta_4 > & = & \frac{1}{N} \left( <L^6> - <L^4> <L^2> \right)
\left( \sum_{\delta=0}^{N-1} \frac{N-\delta}{N} \left[ C(L^2,L^4,\delta) +
C(L^4,L^2,\delta) \right] - 1 \right) \nonumber\\
<\eta_2^2> & = & \frac{1}{N} \left( <L^4> - <L^2>^2 \right)
\left( \sum_{\delta=0}^{N-1} \frac{N-\delta}{N} \left[ 2 C(L^2,L^2,\delta)
- 1 \right] \right)
\end{eqnarray}
and
\begin{equation}
C(X,Y,\delta) = \frac{< (X_i - <X>) (Y_{i+\delta} - <Y>) >}{< (X_i - <X>)
(Y_i - <Y>) >}.
\end{equation}
If $N$ is much bigger than the correlation times of interest then
$C(X,Y,\delta) \simeq 0$ for
$\delta > \bar{\delta}$, with $\bar{\delta} < n$, so we can write
approximately
\begin{eqnarray}
<\eta_2 \eta_4 > & \simeq & \frac{1}{N} \left( <L^6> - <L^4> <L^2> \right)
\left( \tau_{\text{int}}(L^2,L^4) + \tau_{\text{int}}(L^4,L^2) - 1 \right)
\\
<\eta_2^2>  & \simeq & \frac{1}{N} \left( <L^4> - <L^2>^2 \right) (2
\tau_{\text{int}}(L^2,L^2) - 1)
\end{eqnarray}
where we have defined the integrated autocorrelation time
\begin{equation}
\tau_{\text{int}}(X,Y) = \sum_{\delta=0}^{\infty} C(X,Y,\delta).
\end{equation}
In order to correct for the $O(\frac{1}{N})$ bias we note that in
Eq.~(\ref{eq:fluctuationsg4}) the terms linear in the fluctuations do
not contribute to the expectation value. So it is sufficient to
define an ``improved'' estimator $g_4^\star$ as
\begin{eqnarray}
\label{eq:g4star}
g_4^\star = \bar{g}_4 \left[ 1 - 3 \frac{\eta_2^2}{<L^2>^2} +
2 \frac{\eta_2 \eta_4}{<L^2><L^4>} \right] =
g_4 \left[ 1 - 2 \frac{\eta_2}{<L^2>} + \frac{\eta_4}{<L^4>} +
o\left( \frac{1}{N} \right) \right].
\end{eqnarray}
To the order we are working we can substitute the quadratic fluctuations
terms with the ``naive'' estimator for their expectation value, and we
obtain
\begin{eqnarray}
g_4^\star & = &
\frac{\frac{1}{N} \sum_i L_i^4}{\frac{1}{N^2} \sum_{i,j} L_i^2 L_j^2}
\left[1 - \frac{3}{N}
\left( \frac{\frac{1}{N} \sum_i L_i^4}{\frac{1}{N^2} \sum_{i,j} L_i^2
L_j^2} - 1 \right) E\left[2 \tau_{\text{int}}(L^2,L^2)-1\right] + \right.
\nonumber \\
 & & + \left. \frac{2}{N}
\left( \frac{\frac{1}{N} \sum_i L_i^6}{\frac{1}{N^2} \sum_{i,j} L_i^2
L_j^4} - 1 \right) E\left[\tau_{\text{int}}(L^2,L^2) +
\tau_{\text{int}}(L^2,L^4) -1\right]
\right].
\end{eqnarray}
To evaluate the variance of $g_4^\star$ we can write using
Eq.~(\ref{eq:g4star})
\begin{equation}
\sigma^2(g_4^\star) = <(g_4^\star)^2> - g_4^2 = g_4^2
\left[ 4 \frac{<\eta_2^2>}{<L^2>^2} + \frac{<\eta_4^2>}{<L^4>^2} -
4 \frac{<\eta_2 \eta_4>}{<L^2><L^4>} \right].
\end{equation}
We can estimate this quantity to the lowest order with
\begin{eqnarray}
E\left[\sigma^2(g_4^\star)\right] & = &
\left( \frac{\frac{1}{N} \sum_i L_i^4}
{\frac{1}{N^2} \sum_{i,j} L_i^2 L_j^2} \right)^2
\left[
\frac{4}{N}
\left( \frac{\frac{1}{N} \sum_i L_i^4}{\frac{1}{N^2} \sum_{i,j} L_i^2
L_j^2} - 1 \right) E\left[2 \tau_{\text{int}}(L^2,L^2)-1\right] + \right.
\nonumber \\
 & & \left. +
\frac{1}{N}
\left( \frac{\frac{1}{N} \sum_i L_i^8}{\frac{1}{N^2} \sum_{i,j} L_i^4
L_j^4} - 1 \right) E\left[2 \tau_{\text{int}}(L^4,L^4)-1\right] + \right.
\nonumber \\
& & \left. -\frac{4}{N}
\left( \frac{\frac{1}{N} \sum_i L_i^6}{\frac{1}{N^2} \sum_{i,j} L_i^2
L_j^4} - 1 \right) E\left[\tau_{\text{int}}(L^2,L^2) +
\tau_{\text{int}}(L^2,L^4)
-1\right] \right].
\end{eqnarray}

\section{Density of States method}
\label{dsm}
To extract the value of an observable ${\cal O}$ from a monte carlo
simulation we average over the sequence of values $O_i$ generated
by the algorithm. This is nothing else than an approximation of
the path integral formula
\begin{equation}
<{\cal O}>(\beta) =
\frac{\int {\cal D}\psi {\cal O}\left[\psi\right] \exp \left( - \beta
S\left[\psi\right] \right)}
{\int {\cal D}\psi  \exp \left( - \beta S\left[\psi\right] \right)}
\simeq \frac{1}{N} \sum_i O_i.
\end{equation}
Using the fact that we know the probability distribution for the fundamental
fields we can extrapolate the observable's value around the $\beta$ value we
have used in the simulation.
To this effect we note that
\begin{eqnarray}
<{\cal O}>(\beta + \Delta \beta) & = &
\frac{\int {\cal D}\psi {\cal O}\left[\psi\right]
\exp \left( - \Delta \beta S\left[\psi\right] \right)
\exp \left( - \beta S\left[\psi\right] \right)}
{\int {\cal D}\psi  \exp \left( - \Delta \beta S\left[\psi\right] \right)
\exp \left( - \beta S\left[\psi\right] \right)} =
\frac{<{\cal O} \exp \left( - \Delta \beta S \right) >(\beta)}
{<\exp \left( - \Delta \beta S \right) >(\beta) } \\
 <{\cal O}>(\beta + \Delta \beta) & \simeq &
\frac{\sum_i O_i \exp \left( - \Delta \beta S_i \right)}
{\sum_i \exp \left( - \Delta \beta S_i \right)},
\label{eq:resum}
\end{eqnarray}
so it is sufficient to measure the action $S_i$ correlated with
each $O_i$ value, and resum each measure according to
Eq.~(\ref{eq:resum}).
The ``resummed'' value will be reliable only for small $\Delta \beta$,because
we have a good statistical sampling of action distribution only in a small
range around the mean value, so we must do a preliminary determination of
the allowed resummation range. During our simulation we have used
essentially the same method as Alves et al.
in~\cite{alves:berg:sanielevici}: we plot histograms of the energy
distribution, we set a minimum and maximum energy by imposing that at least
$2.5\%$ of the energy distribution is present below the minimum and above
the maximum, and we translate this energy range in a $\beta$ range by
finding the $\beta$ shifts that via resummation give the found energy limits.

If the resumming ranges of different simulations overlap we can ``patch'' the
values coming from different extrapolations to obtain a more accurate
estimate.

The determination of an unbiased estimator for the resummed Binder cumulant
and its variance can be obtained with the same method as in Appendix
\ref{binder}.
We only note that there are some complications when we combine together
different extrapolations, because in this case to cancel the bias we must
use an improved estimator for the variances, depending on many-point
correlations which can be estimated reliably only with very high statistics.

%
%	tables
%
\begin{table}
\begin{tabular}{ccdcdd}
$N_\sigma$ & $N_\tau$ & $\beta$ & $N_{\text{meas}}$ & $\tau_{\text{int}}$
& \text{Rew.\ range}\\
\hline
4 & 2 & 1.340000 & 40960 & 32.8 & 1.28 - 1.40 \\
4 & 2 & 1.360000 & 40960 & 51.3 & 1.30 - 1.42 \\
4 & 2 & 1.380000 & 40960 & 105.7 & 1.32 - 1.44 \\
4 & 2 & 1.400000 & 16384 & 79.4 & 1.34 - 1.46 \\
\hline
8 & 2 & 1.340000 & 16384 & 55.0 & 1.31 - 1.37\\
8 & 2 & 1.360000 & 38912 & 154.4 & 1.33 - 1.39 \\
8 & 2 & 1.380000 & 38912 & 314.9 & 1.35 - 1.41\\
8 & 2 & 1.400000 & 38912 & 1264.8 & 1.37 - 1.43\\
\hline
12 & 2 & 1.360000 & 36864 & 142.9 & 1.34 - 1.37 \\
12 & 2 & 1.380000 & 45056 & 407.1 & 1.36 - 1.39 \\
12 & 2 & 1.400000 & 49152 & 3349.7 & 1.39 - 1.41 \\
\end{tabular}
\caption{Run parameters}
\label{sim:rp1}
\end{table}

\begin{table}
\begin{tabular}{ccdcdd}
$N_\sigma$ & $N_\tau$ & $\beta$ & $N_{\text{meas}}$ & $\tau_{\text{int}}$ &
\text{ Rew.\ range}\\
\hline
6 & 3 & 1.5600 & 32768 & 3.0 & 1.52 - 1.60 \\
6 & 3 & 1.5700 & 32768 & 3.4 & 1.53 - 1.61 \\
6 & 3 & 1.5800 & 32768 & 4.2 & 1.54 - 1.62 \\
6 & 3 & 1.5900 & 32768 & 4.9 & 1.55 - 1.63 \\
6 & 3 & 1.6000 & 32768 & 6.7 & 1.56 - 1.64 \\
6 & 3 & 1.6100 & 32768 & 7.9 & 1.57 - 1.65 \\
6 & 3 & 1.6200 & 32768 & 10.6 & 1.58 - 1.66 \\
6 & 3 & 1.6300 & 32768 & 15.3 & 1.59 - 1.67 \\
\hline
10 & 3 & 1.5600 & 30720 & 5.9 & 1.54 - 1.58 \\
10 & 3 & 1.5700 & 30720 & 8.9 & 1.55 - 1.59 \\
10 & 3 & 1.5800 & 30720 & 15.9 & 1.56 - 1.60 \\
10 & 3 & 1.5900 & 30720 & 22.3 & 1.57 - 1.61 \\
10 & 3 & 1.6000 & 30720 & 53.5 & 1.58 - 1.62 \\
10 & 3 & 1.6100 & 30720 & 103.6 & 1.59 - 1.63 \\
10 & 3 & 1.6200 & 30720 & 138.7 & 1.60 - 1.64 \\
10 & 3 & 1.6300 & 30720 & 327.6 & 1.61 - 1.65 \\
\hline
14 & 3 & 1.5600 & 30336 & 6.5 & 1.551 - 1.569\\
14 & 3 & 1.5700 & 65536 & 10.3 & 1.560 - 1.578\\
14 & 3 & 1.5800 & 65536 & 17.6 & 1.572 - 1.590\\
14 & 3 & 1.5900 & 65536 & 33.6 & 1.579 - 1.599\\
14 & 3 & 1.6000 & 65536 & 84.8 & 1.590 - 1.612\\
14 & 3 & 1.6100 & 30336 & 206.9 & 1.602 - 1.621\\
14 & 3 & 1.6200 & 30336 & 902.2 & 1.612 - 1.629\\
14 & 3 & 1.6300 & 30336 & 3.9 & 1.620 - 1.639\\
\hline
18 & 3 & 1.5800 & 22464 & 12.3 & 1.572 - 1.587 \\
18 & 3 & 1.5900 & 22464 & 22.1 & 1.584 - 1.597 \\
18 & 3 & 1.6000 & 22464 & 81.1 & 1.592 - 1.606 \\
18 & 3 & 1.6100 & 22464 & 563.7 & 1.602 - 1.617 \\
\end{tabular}
\caption{Run parameters}
\label{sim:rp2}
\end{table}

\begin{table}
\begin{tabular}{ccdcdd}
$N_\sigma$ & $N_\tau$ & $\beta$ & $N_{\text{meas}}$ & $\tau_{\text{int}}$ &
\text{ Rew.\ range}\\
\hline
8 & 4 & 1.6500 & 65536 & 3.2 & 1.625 - 1.675 \\
8 & 4 & 1.6700 & 65536 & 4.8 & 1.646 - 1.694 \\
8 & 4 & 1.6900 & 65536 & 7.6 & 1.665 - 1.716 \\
8 & 4 & 1.6950 & 131072 & 8.7 & 1.670 - 1.721 \\
8 & 4 & 1.7050 & 131072 & 11.0 & 1.680 - 1.731 \\
8 & 4 & 1.7100 & 196608 & 12.9 & 1.686 - 1.736 \\
8 & 4 & 1.7200 & 262144 & 18.5 & 1.694 - 1.746 \\
8 & 4 & 1.7300 & 131072 & 23.0 & 1.704 - 1.757 \\
\hline
12 & 4 & 1.6500 & 4096 & 3.5 & 1.637 - 1.665 \\
12 & 4 & 1.6700 & 4096 & 6.6 & 1.657 - 1.684 \\
12 & 4 & 1.6900 & 69632 & 12.5 & 1.676 - 1.704 \\
12 & 4 & 1.6950 & 131072 & 15.5 & 1.681 - 1.709 \\
12 & 4 & 1.7000 & 65536 & 22.7 & 1.686 - 1.714 \\
12 & 4 & 1.7050 & 131072 & 29.1 & 1.691 - 1.719 \\
12 & 4 & 1.7100 & 135168 & 50.5 & 1.704 - 1.724 \\
12 & 4 & 1.7200 & 131072 & 84.4 & 1.714 - 1.734 \\
12 & 4 & 1.7300 & 131072 & 200.7 & 1.717 - 1.744 \\
12 & 4 & 1.7400 & 131072 & 351.8 & 1.726 - 1.755 \\
\hline
16 & 4 & 1.6900 & 28672 & 16.3 & 1.681 - 1.699 \\
16 & 4 & 1.6950 & 57344 & 25.4 & 1.686 - 1.704 \\
16 & 4 & 1.7000 & 57344 & 52.1 & 1.691 - 1.707 \\
16 & 4 & 1.7050 & 40960 & 55.4 & 1.696 - 1.714 \\
16 & 4 & 1.7100 & 12288 & 71.2 & 1.701 - 1.719 \\
\end{tabular}
\caption{Run parameters (continued) }
\label{sim:rp3}
\end{table}

\begin{table}
\begin{tabular}{ccdcdd}
$N_\sigma$ & $N_\tau$ & $\beta$ & $N_{\text{meas}}$ & $\tau_{\text{int}}$ &
\text{ Rew.\ range}\\
\hline
8 & 5 & 1.7500 & 16384 & 6.7 & 1.727 - 1.772 \\
8 & 5 & 1.7600 & 40960 & 7.4 & 1.738 - 1.785 \\
8 & 5 & 1.7650 & 40960 & 8.4 & 1.741 - 1.791 \\
8 & 5 & 1.7700 & 40960 & 9.0 & 1.745 - 1.796\\
8 & 5 & 1.7750 & 40960 & 10.0 & 1.750 - 1.801 \\
8 & 5 & 1.7800 & 57344 & 11.3 & 1.755 - 1.806\\
8 & 5 & 1.7850 & 40960 & 12.0 & 1.759 - 1.811\\
8 & 5 & 1.7900 & 40960 & 12.4 & 1.762 - 1.815\\
8 & 5 & 1.7950 & 40960 & 13.1 & 1.770 - 1.822\\
\hline
12 & 5 & 1.7600 & 39936 & 14.0 & 1.746 - 1.774 \\
12 & 5 & 1.7650 & 39936 & 18.2 & 1.751 - 1.777 \\
12 & 5 & 1.7700 & 39936 & 22.9 & 1.756 - 1.783 \\
12 & 5 & 1.7750 & 39936 & 28.5 & 1.761 - 1.789 \\
12 & 5 & 1.7800 & 23552 & 27.5 & 1.766 - 1.792 \\
12 & 5 & 1.7850 & 23552 & 43.1 & 1.771 - 1.798 \\
12 & 5 & 1.7900 & 23552 & 55.8 & 1.776 - 1.804 \\
12 & 5 & 1.7950 & 23552 & 51.6 & 1.780 - 1.808 \\
\hline
16 & 5 & 1.7600 & 28672 & 24.0 & 1.751 - 1.767 \\
16 & 5 & 1.7650 & 28672 & 24.2 & 1.756 - 1.774 \\
16 & 5 & 1.7700 & 28672 & 35.6 & 1.761 - 1.779 \\
16 & 5 & 1.7750 & 28672 & 46.4 & 1.768 - 1.784 \\
\hline
20 & 5 & 1.7670 & 18432 & 52.4 & 1.761 - 1.773 \\
20 & 5 & 1.7690 & 18432 & 58.5 & 1.763 - 1.774 \\
\end{tabular}
\caption{Run parameters (continued)}
\label{sim:rp4}
\end{table}

\begin{table}
\begin{tabular}{ccdcdd}
$N_\sigma$ & $N_\tau$ & $\beta$ & $N_{\text{meas}}$ & $\tau_{\text{int}}$ &
\text{ Rew.\ range}\\
\hline
8 & 6 & 1.7750 & 32768 & 3.9 & 1.754 - 1.799 \\
8 & 6 & 1.8000 & 32768 & 5.4 & 1.7775 - 1.824 \\
8 & 6 & 1.8250 & 32768 & 7.5 & 1.800 - 1.851 \\
8 & 6 & 1.8750 & 32768 & 11.2 & 1.850 - 1.902 \\
\hline
12 & 6 & 1.7750 & 49152 & 5.6 & 1.762 - 1.7875 \\
12 & 6 & 1.8000 & 49152 & 10.1 & 1.787 - 1.812 \\
12 & 6 & 1.8125 & 16384 & 14.7 & 1.799 - 1.826 \\
12 & 6 & 1.8250 & 49152 & 19.5 & 1.811 - 1.837 \\
12 & 6 & 1.8375 & 16384 & 30.6 & 1.824 - 1.850 \\
12 & 6 & 1.8750 & 49152 & 87.9 & 1.861 - 1.889 \\
\hline
16 & 6 & 1.7750 & 28672 & 7.5 & 1.767 - 1.783 \\
16 & 6 & 1.8000 & 28672 & 12.0 & 1.792 - 1.809 \\
16 & 6 & 1.8125 & 16384 & 21.9 & 1.805 - 1.820 \\
16 & 6 & 1.8250 & 28672 & 30.1 & 1.817 - 1.832 \\
16 & 6 & 1.8375 & 16384 & 51.2 & 1.829 - 1.846 \\
16 & 6 & 1.8750 & 28672 & 466.3 & 1.886 - 1.884 \\
\hline
20 & 6 & 1.8200 & 16384 & 30.3 & 1.814 - 1.825 \\
20 & 6 & 1.8250 & 18920 & 55.0 & 1.819 - 1.831 \\
20 & 6 & 1.8300 & 16384 & 57.6 & 1.825 - 1.836 \\
\hline
24 & 6 & 1.8220 & 12288 & 28.5 & 1.818 - 1.826 \\
24 & 6 & 1.8250 & 27008 & 62.9 & 1.821 - 1.829 \\
24 & 6 & 1.8280 & 27008 & 80.2 & 1.823 - 1.832 \\
\end{tabular}
\caption{Run parameters (continued)}
\label{sim:rp5}
\end{table}

\begin{table}
\begin{tabular}{ccdcdd}
$N_\sigma$ & $N_\tau$ & $\beta$ & $N_{\text{meas}}$ & $\tau_{\text{int}}$ &
\text{ Rew.\ range}\\
\hline
14 & 7 & 1.8700 & 9216 & 31.3 & 1.861 - 1.880 \\
14 & 7 & 1.8750 & 11600 & 27.2 & 1.867 - 1.884 \\
14 & 7 & 1.8800 & 12288 & 23.5 & 1.872 - 1.889 \\
14 & 7 & 1.8850 & 9216 & 32.0 & 1.878 - 1.892 \\
\hline
20 & 7 & 1.8725 & 15904 & 40.7 & 1.867 - 1.877  \\
20 & 7 & 1.8750 & 17216 & 47.8 & 1.870 - 1.880 \\
20 & 7 & 1.8775 & 18400 & 58.3 & 1.871 - 1.884 \\
20 & 7 & 1.8800 & 14720 & 77.4 & 1.874 - 1.885 \\
\hline
16 & 8 & 1.9000 & 13312 & 19.0 & 1.891 - 1.909 \\
16 & 8 & 1.9150 & 11536 & 34.0 & 1.907 - 1.922 \\
16 & 8 & 1.9200 & 13312 & 46.2 & 1.911 - 1.929 \\
16 & 8 & 1.9250 & 16384 & 39.2 & 1.918 - 1.932 \\
\hline
24 & 8 & 1.9000 & 18432 & 33.1 & 1.896 - 1.904 \\
24 & 8 & 1.9150 & 29184 & 54.5 & 1.911 - 1.919 \\
24 & 8 & 1.9180 & 10144 & 27.4 & 1.914 - 1.922 \\
24 & 8 & 1.9200 & 20736 & 54.9 & 1.916 - 1.924 \\
24 & 8 & 1.9220 & 10144 & 41.4 & 1.918 - 1.926 \\
24 & 8 & 1.9250 & 6144 & 54.9 & 1.921 - 1.929 \\
\end{tabular}
\caption{Run parameters (continued)}
\label{sim:rp6}
\end{table}

\begin{table}
\begin{tabular}{ccdccdd}
$N_\sigma$ & $N_\tau$ & $\beta$ & $N_{\text{meas}}$ & $\tau_{\text{int}}$ & $
U_{1\times 1}$ &
$ U_{1\times 2} $ \\
\hline
12 & 12 & 1.375000 & 384 & 1.0 & 0.51106(11) & 0.25133(14) \\
12 & 12 & 1.385000 & 384 & 1.0 & 0.51488(13) & 0.25561(20) \\
\hline
14 & 14 & 1.590000 & 384 & 1.0 & 0.596484(95) & 0.35759(14)\\
14 & 14 & 1.595000 & 384 & 1.0 & 0.598358(84) & 0.36011(12)\\
14 & 14 & 1.600000 & 384 & 1.0 & 0.600174(94) & 0.36261(15)\\
\hline
16 & 16 & 1.690000 & 384 & 1.0 & 0.631531(63) & 0.40691(11) \\
16 & 16 & 1.695000 & 384 & 1.1 & 0.633037(61) & 0.409088(95) \\
16 & 16 & 1.700000 & 384 & 1.0 & 0.634699(72) & 0.41150(12) \\
16 & 16 & 1.705000 & 384 & 1.0 & 0.636203(78) & 0.41371(10) \\
\hline
20 & 20 & 1.769000 & 2304 & 1.0 & 0.654686(17) & 0.440752(27)\\
20 & 20 & 1.770000 & 2304 & 1.0 & 0.654941(13) & 0.441124(22)\\
\hline
20 & 20 & 1.827000 & 2048 & 1.0 & 0.669217(16) & 0.462206(26)\\
20 & 20 & 1.828500 & 3072 & 1.0 & 0.669557(10) & 0.462711(17)\\
20 & 20 & 1.830000 & 2048 & 1.0 & 0.669923(13) & 0.463250(22)\\
\hline
20 & 20 & 1.872500 & 1536 & 1.2 & 0.679420(18) & 0.477325(29)\\
20 & 20 & 1.877500 & 1536 & 1.0 & 0.680491(14) & 0.478897(24)\\
\hline
20 & 20 & 1.915000 & 1536 & 1.0 & 0.688180(16) & 0.490325(27)\\
20 & 20 & 1.925000 & 1536 & 1.0 & 0.690137(15) & 0.493229(24)\\
\end{tabular}
\caption{Run parameters on symmetric lattices, and expectation values of
plaquettes}
\label{sim:rp7}
\end{table}

\begin{table}
\begin{tabular}[t]{cccd}
$N_\tau$ & $N_\sigma$ & $N_{\sigma^\prime}$ & $\beta_c$ \\
\hline
2 & 4 & 8 & 1.380(4) \\
2 & 4 & 12 & 1.380(2) \\
2 & 8 & 12 & 1.380(4) \\
\hline
3 & 6 & 10 & 1.595(1) \\
3 & 6 & 14 & 1.5955(8) \\
3 & 6 & 18 & 1.5958(5) \\
3 & 10 & 14 & 1.5956(12) \\
3 & 10 & 18 & 1.5960(7) \\
3 & 14 & 18 & 1.5963(15) \\
\hline
4 & 8 & 12 & 1.6984(7) \\
4 & 8 & 16 & 1.6987(5) \\
4 & 12 & 16 & 1.699(1) \\
\hline
5 & 8 & 12 & 1.7669(12) \\
5 & 8 & 16 & 1.7678(8) \\
5 & 8 & 20 & 1.7682(7) \\
5 & 12 & 16 & 1.7684(15) \\
5 & 12 & 20 & 1.769(12) \\
5 & 16 & 20 & 1.769(3) \\
\hline
6 & 8 & 12 & 1.812(4) \\
6 & 8 & 16 & 1.817(2) \\
6 & 8 & 20 & 1.822(2) \\
6 & 8 & 24 & 1.824(1) \\
6 & 12 & 16 & 1.823(5) \\
6 & 12 & 20 & 1.828(2) \\
6 & 12 & 24 & 1.825(1) \\
6 & 16 & 20 & 1.830(4) \\
6 & 16 & 24 & 1.825(2) \\
6 & 20 & 24 & 1.824(3) \\
\hline
7 & 14 & 20 & 1.8747(20) \\
\hline
8 & 16 & 24 & 1.920(4) \\
\end{tabular}
\caption{Crossings for $g_4$ curves}
\label{fss:crossings}
\end{table}

\begin{table}
\begin{tabular}{cdd}
$N_\tau$ & $\beta_c$ & $T_c / \Lambda_{\text{I}}$ \\
\hline
2 & 1.380(4) & 8.81(8) \\
3 & 1.59624(13) & 9.889(3) \\
4 & 1.699(1) & 9.526(30) \\
5 & 1.76948(3) & 9.0562(7) \\
6 & 1.8287(11) & 8.73(3) \\
7 & 1.8747(30) & 8.38(6) \\
8 & 1.920(5)   & 8.2(1) \\
\end{tabular}
\caption{Critical temperature by two loop asymptotic scaling formula}
\label{fss:tc}
\end{table}

\begin{table}
\begin{tabular}{cddddd}
$N_\tau$ & $\beta_c$ & $T_c / \Lambda$ & $A$ & $\eta$ & $\chi^2$ \\
\hline
3 & 1.588(2) & 9.70(5) & 0.76(4) & 0.35(2) & 5.4 \\
4 & 1.687(2) & 9.25(3) & 0.66(4) & 0.40(2) & 15.3 \\
5 & 1.759(3) & 8.82(5) & 0.47(2) & 0.38(2) & 4.0 \\
\end{tabular}
\caption{Results of the fit}
\label{fittab}
\end{table}

\begin{table}
\begin{tabular}{cddd}
$N_\tau$ & $\beta_I$ & $\beta_L$ & $\Lambda_I / \Lambda_L$ \\
\hline
2 & 1.380(4) & 1.8800(30) & 3.599(46)\\
3 & 1.59624(13) & 2.1768(30) & 4.470(35)\\
4 & 1.699(1) & 2.2986(6) & 4.713(14)\\
5 & 1.76948(3) & 2.3726(45) & 4.766(56)\\
6 & 1.8287(11) & 2.4265(30) & 4.709(39)\\
8 & 1.920(5)  & 2.5115(40) & 4.644(77)\\
\end{tabular}
\caption{$\Lambda$ ratios from asymptotic formula}
\label{lambda_ratios}
\end{table}

\begin{table}
\begin{tabular}{cdddd}
$N_\tau$ & $\alpha_0$ & $\alpha_v$ & $T/\Lambda_L$ & $T/\Lambda_v$ \\
\hline
2 & 0.1693(3) & 0.2466(6)& 29.7(2) & 2.27(2)\\
3 & 0.1462(2) & 0.2039(3) & 41.4(3) & 2.89(2)\\
4 & 0.13848(4) & 0.19022(7) & 42.1(1) & 2.848(4)\\
5 & 0.1342(2) & 0.1828(3) & 40.6(5) & 2.69(2)\\
6 & 0.1312(2) & 0.1776(3) & 38.7(3) & 2.54(2)\\
8 & 0.1267(2) & 0.1700(3) & 36.0(4) & 2.32(2)\\
16 & 0.1162(4) & 0.1526(6) & 32.0(8) & 1.97(4)\\
\end{tabular}
\caption{Perturbative $\alpha_v$: Wilson case}
\label{lepage:TcW}
\end{table}

\begin{table}
\begin{tabular}{cdddd}
$N_\tau$ & $\alpha_0$ & $\alpha_v$ & $T/\Lambda_I$ & $T/\Lambda_v$ \\
\hline
2 & 0.2306(7) & 0.286(1) & 8.81(8) & 1.49(1) \\
3 & 0.19941(2) & 0.24085(3) & 9.889(3)  & 1.627(6) \\
4 & 0.1873(1) & 0.2239(1) & 9.526(30)  & 1.549(2) \\
5 & 0.179889(3) & 0.213616(4) & 9.0562(7) & 1.461(1) \\
6 & 0.1741(1) & 0.2057(1) & 8.73(3) & 1.398(2) \\
7 & 0.1698(3) & 0.1998(4) & 8.38(6) & 1.34(1) \\
8 & 0.1658(4) & 0.1944(5) & 8.2(1) & 1.30(1) \\
\end{tabular}
\caption{Perturbative $\alpha_v$: Symanzik case}
\label{lepage:TcS}
\end{table}

\begin{table}
\begin{tabular}{lllll}
$\beta$ & $\beta_1$ & $T_1/\Lambda_1$ & $\beta_2$ & $T_2/\Lambda_2$ \\
\hline
1.380(4) & 1.1280(35) & 1.832(15) & 1.331(30) & 2.953(21) \\
1.59624(13) & 1.36942(16) & 2.15542(83) & 1.55480(16) & 3.3651(13) \\
1.699(1) & 1.5026(13) & 2.2247(73) & 1.6876(13) & 3.485(12) \\
1.76948(3) & 1.59156(3) & 2.2072(2) & 1.77784(3) & 3.4781(3) \\
1.8287(11) & 1.6628(13) & 2.1876(70) & 1.850(1) & 3.462(11) \\
1.8747(30) & 1.7163(34) & 2.136(18) & 1.9041(34) & 3.391(29) \\
1.920(5) & 1.7674(56) & 2.119(29) & 1.9557(56) & 3.371(47)\\
\end{tabular}
\caption{Data for the effective schemes}
\label{eff:data}
\end{table}
\clearpage
\clearpage
\clearpage
\begin{figure}
\protect{
\vskip 10truecm
\includegraphics{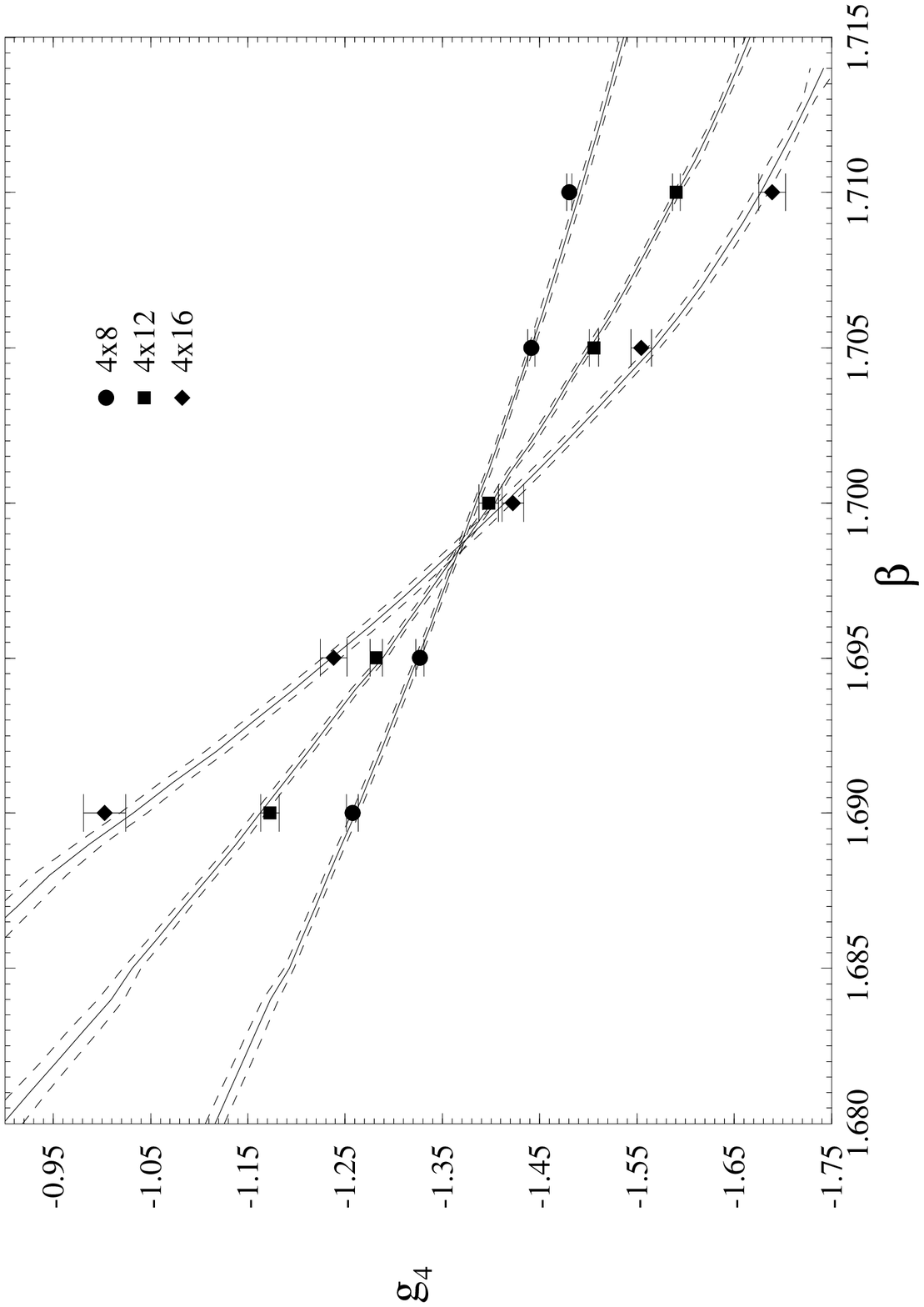}
}
\vskip 1.in
\caption{Binder cumulant: lattices with $N_\tau = 4$}
\label{fig3}
\end{figure}
\clearpage
\begin{figure}
\protect{
\vskip 10truecm
\includegraphics{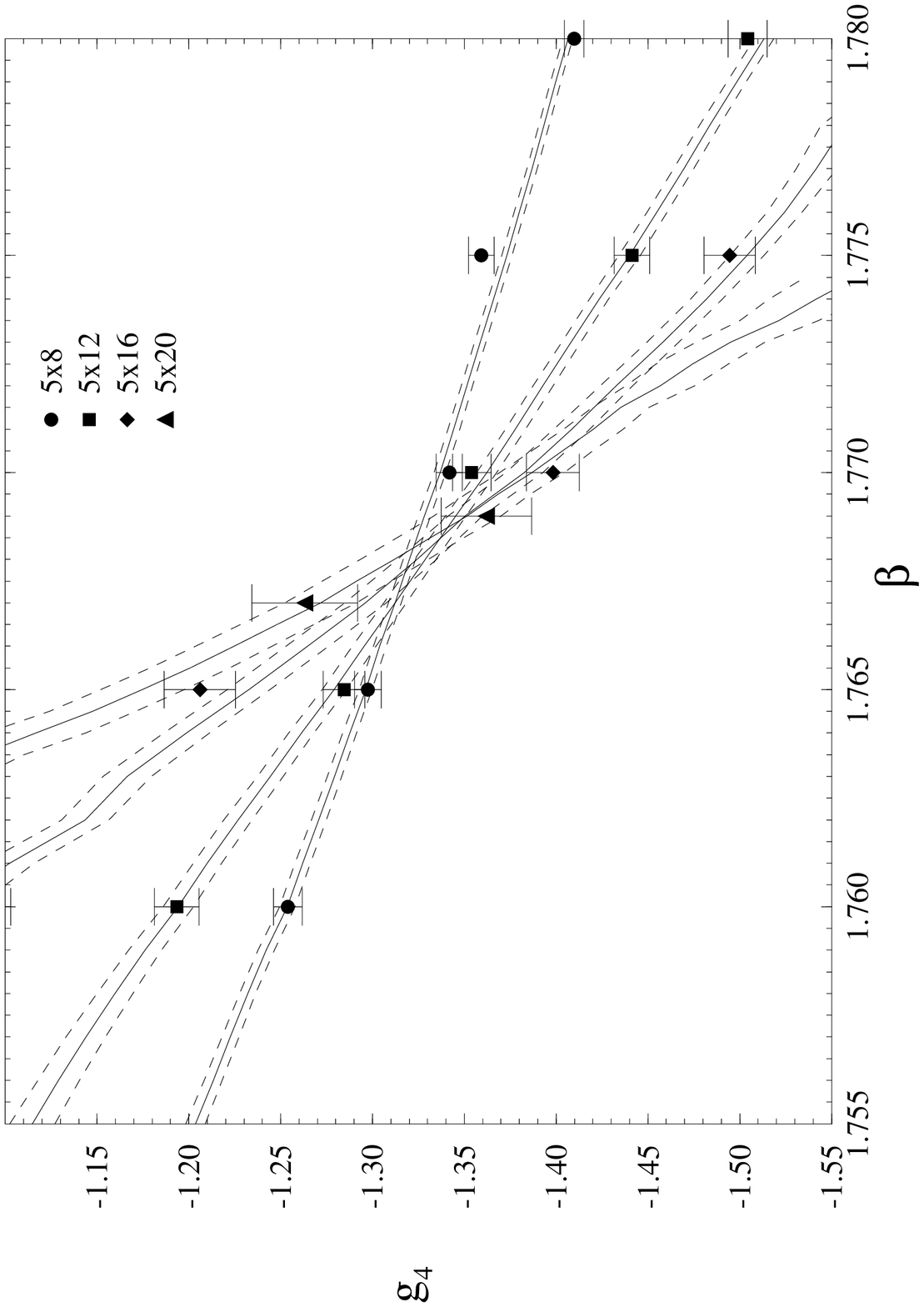}
}
\vskip 1.in
\caption{Binder cumulant: lattices with $N_\tau = 5$}
\label{fig4}
\end{figure}
\clearpage
\begin{figure}
\protect{
\vskip 10truecm
\includegraphics{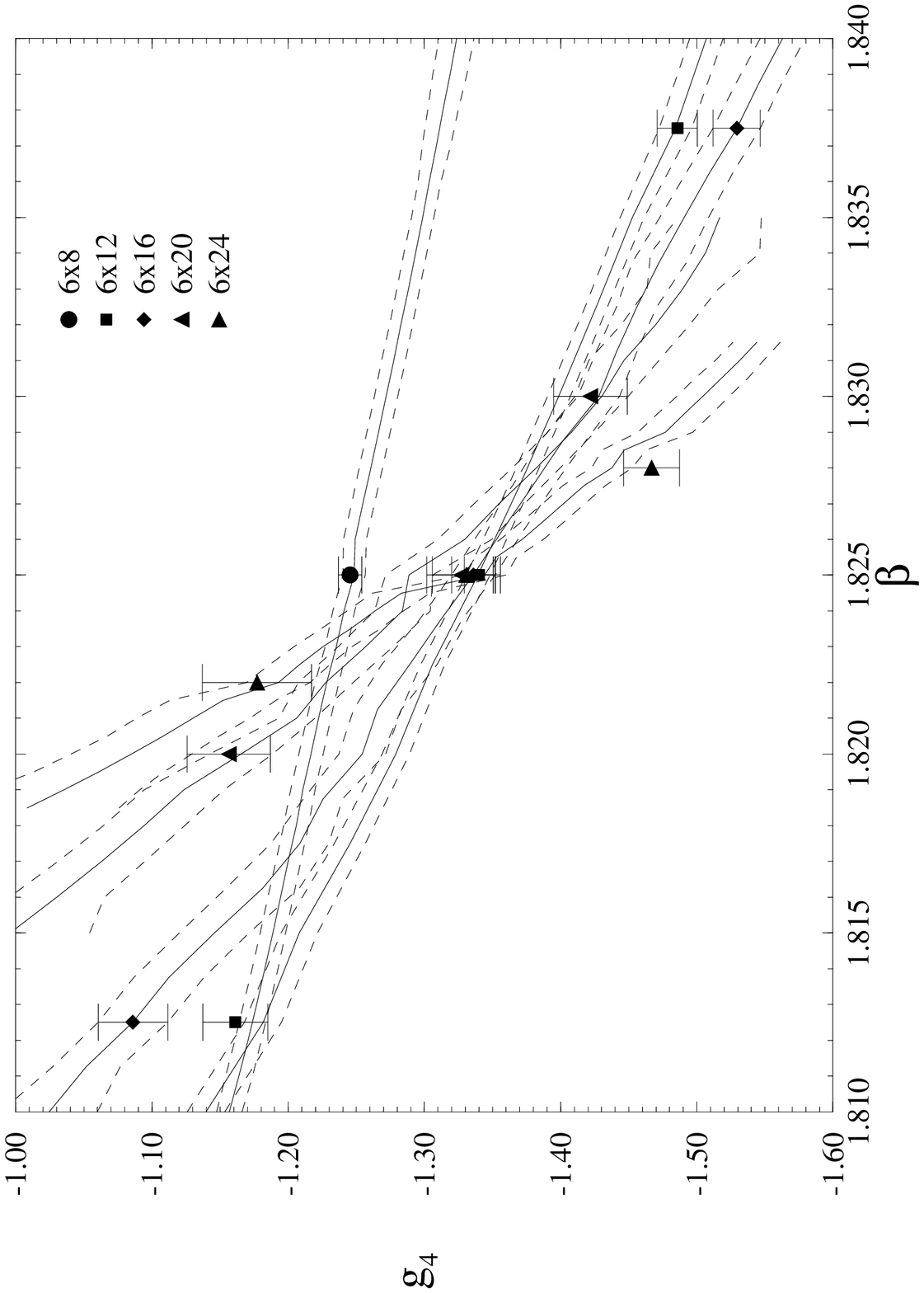}
}
\vskip 1.in
\caption{Binder cumulant: lattices with $N_\tau = 6$}
\label{fig5}
\end{figure}
\clearpage
\begin{figure}
\protect{
\vskip 10truecm
\includegraphics{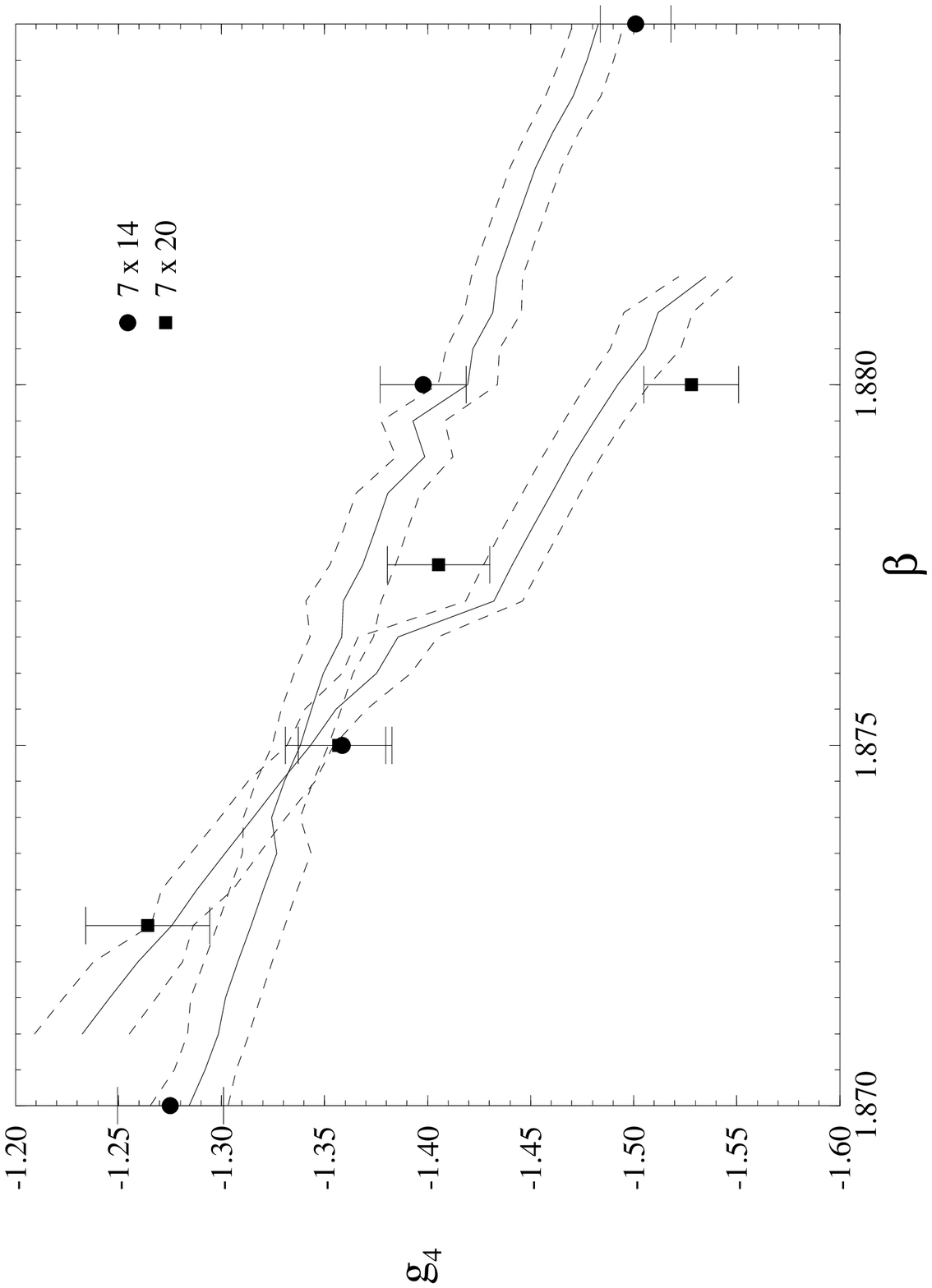}
}
\vskip 1.in
\caption{Binder cumulant: lattices with $N_\tau = 7$}
\label{fig6}
\end{figure}
\clearpage
\begin{figure}
\protect{
\vskip 10truecm
\includegraphics{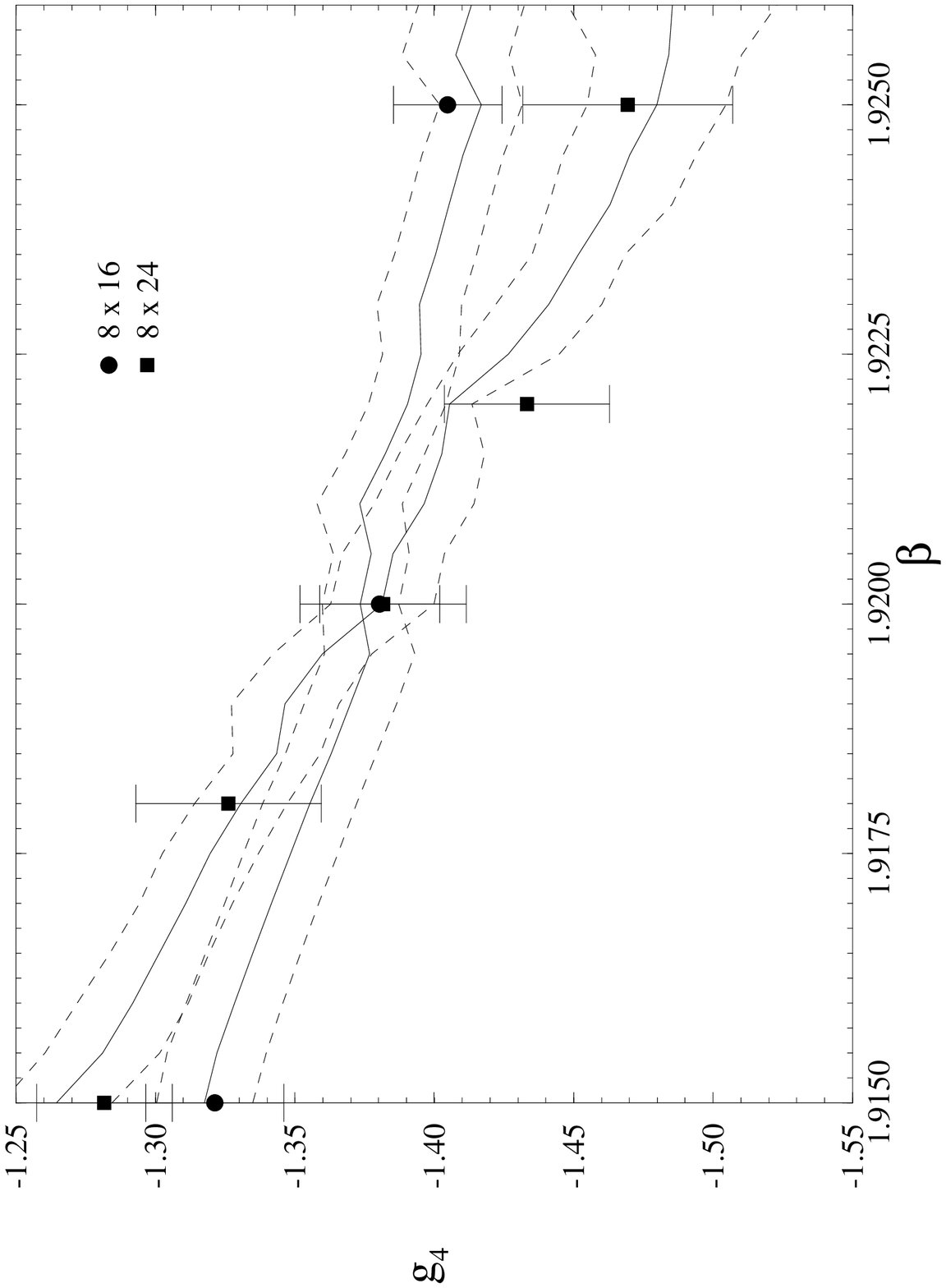}
}
\vskip 1.in
\caption{Binder cumulant: lattices with $N_\tau = 8$}
\label{fig7}
\end{figure}
\clearpage
\begin{figure}
\protect{
\vskip 10truecm
\includegraphics{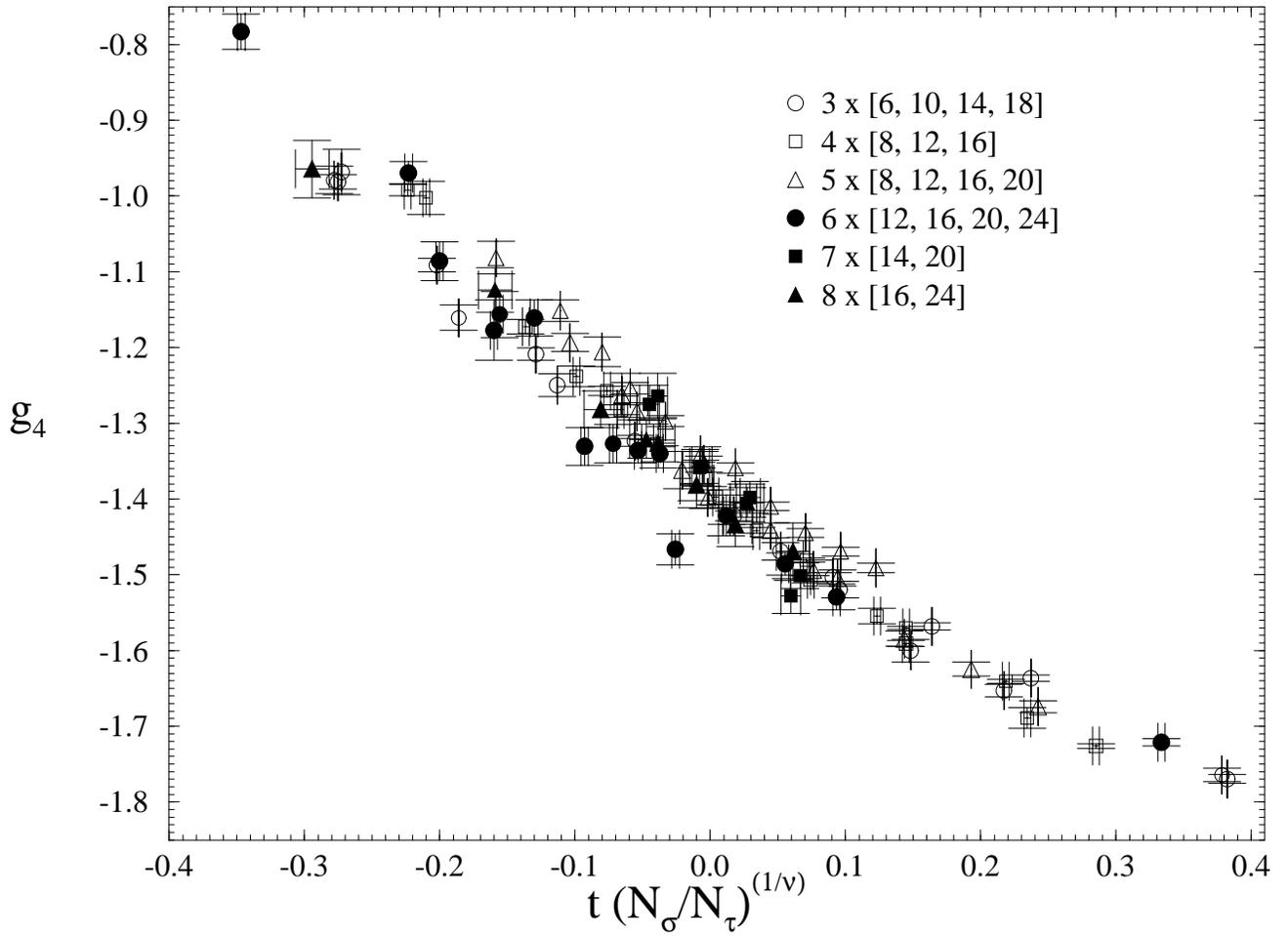}
}
\vskip 1.in
\caption{Universal scaling behavior of Binder cumulant}
\label{fig8}
\end{figure}
\clearpage
\begin{figure}
\protect{
\vskip 10truecm
\includegraphics{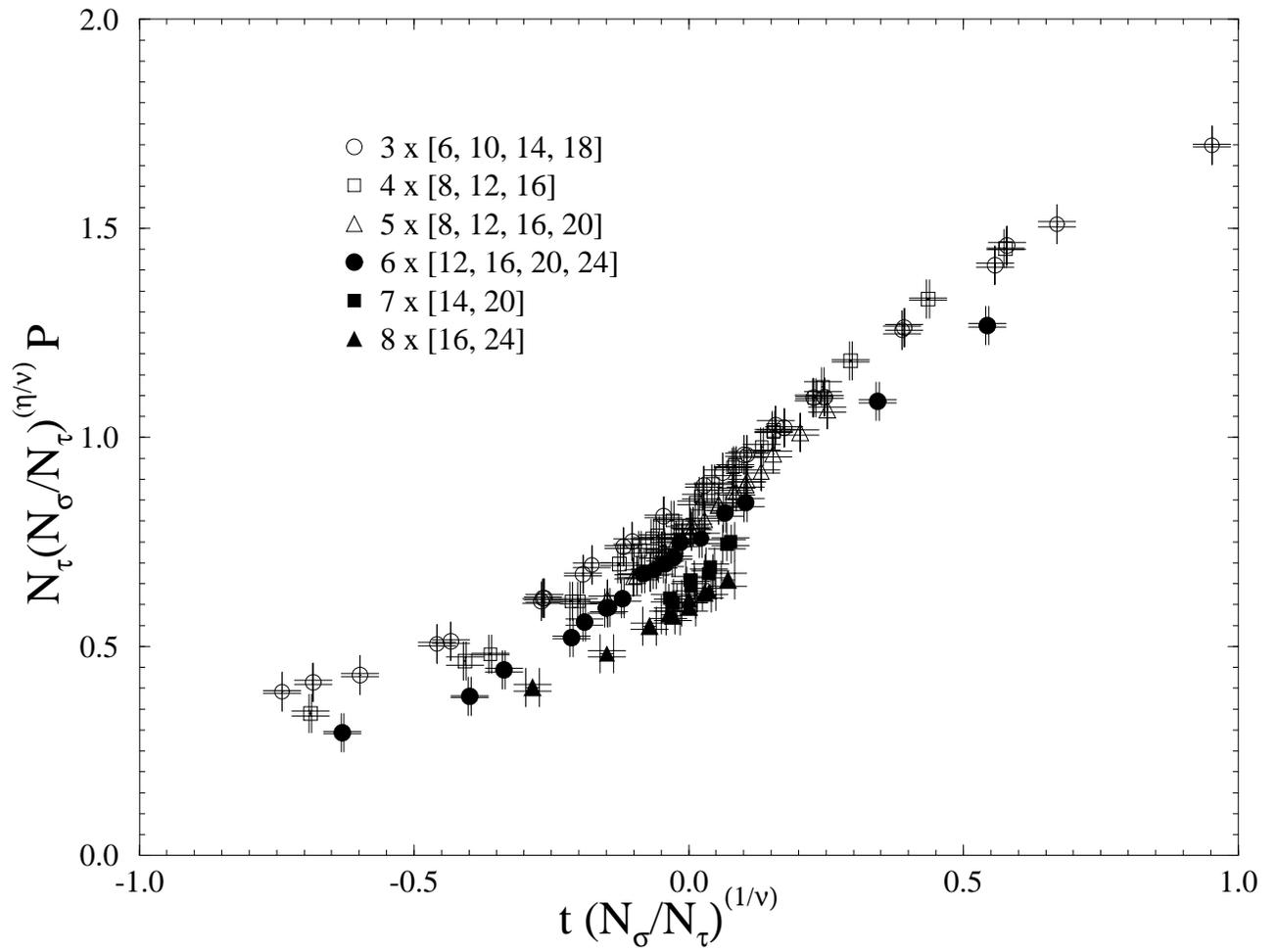}
}
\vskip 1.in
\caption{Universal scaling behavior of Polyakov loop}
\label{fig9}
\end{figure}
\clearpage
\begin{figure}
\protect{
\vskip 10truecm
\includegraphics{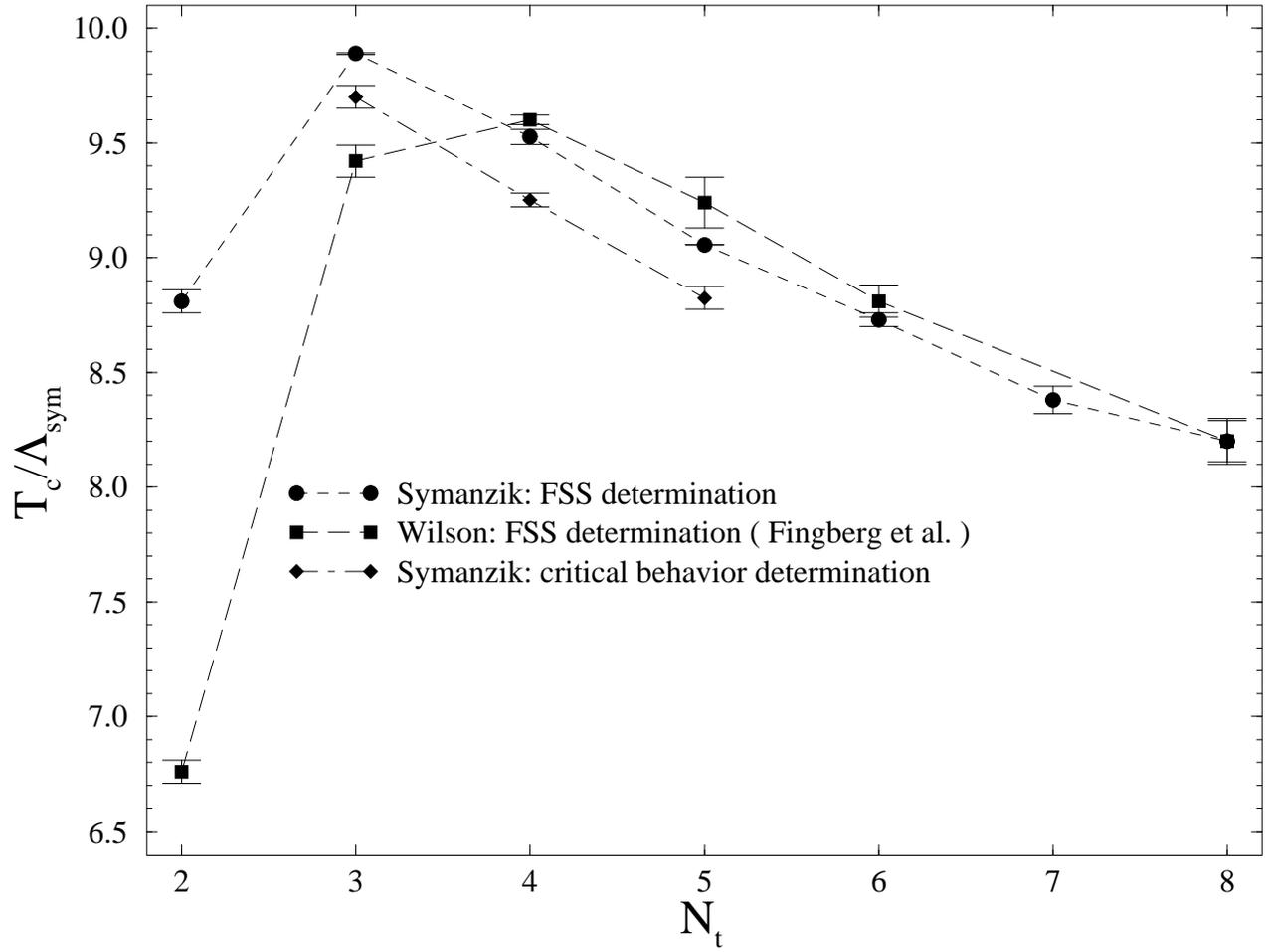}
}
\vskip 1.in
\caption{Plot of the $T_c / \Lambda$ ratio by use of the ``bare'' scheme}
\label{fig10}
\end{figure}
\clearpage
\begin{figure}
\protect{
\vskip 10truecm
\includegraphics{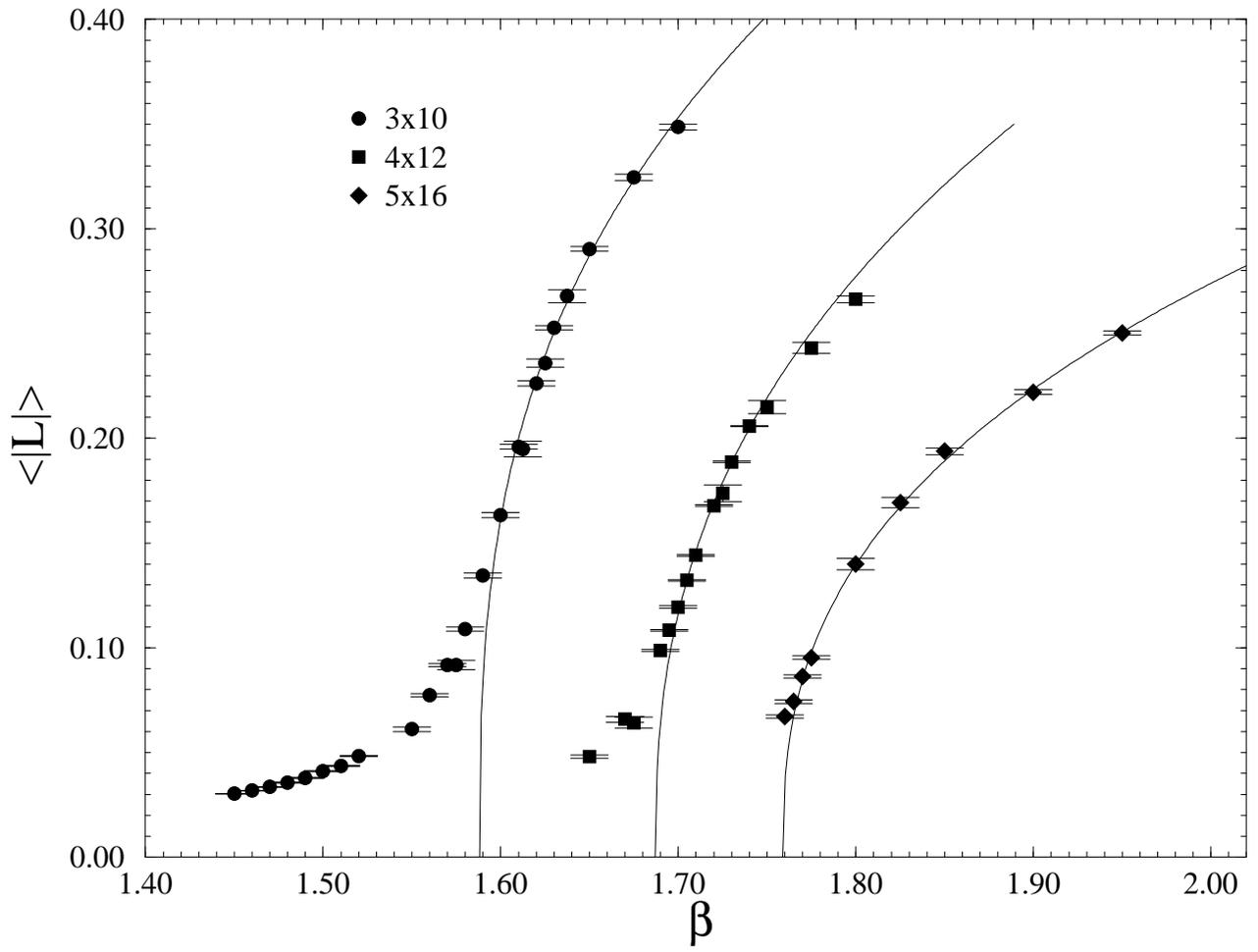}
}
\vskip 1.in
\caption{Critical behavior of Polyakov line}
\label{fig11}
\end{figure}
\clearpage
\begin{figure}
\protect{
\vskip 10truecm
\includegraphics{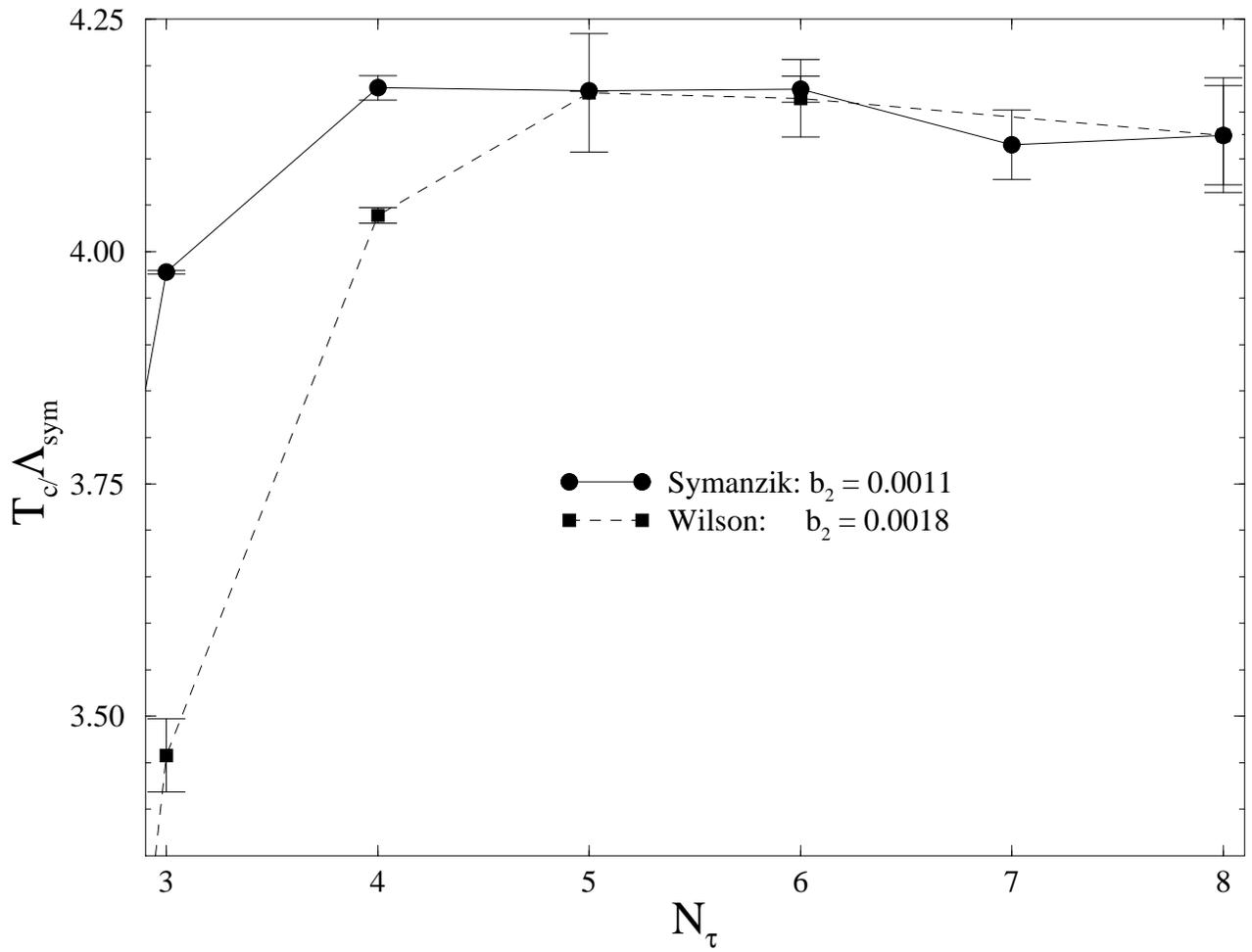}
}
\vskip 1.in
\caption{``Bare'' scheme corrected with the first sub-asymptotic
coefficient of $\beta$ function}
\label{fig12}
\end{figure}
\clearpage
\begin{figure}
\protect{
\vskip 10truecm
\includegraphics{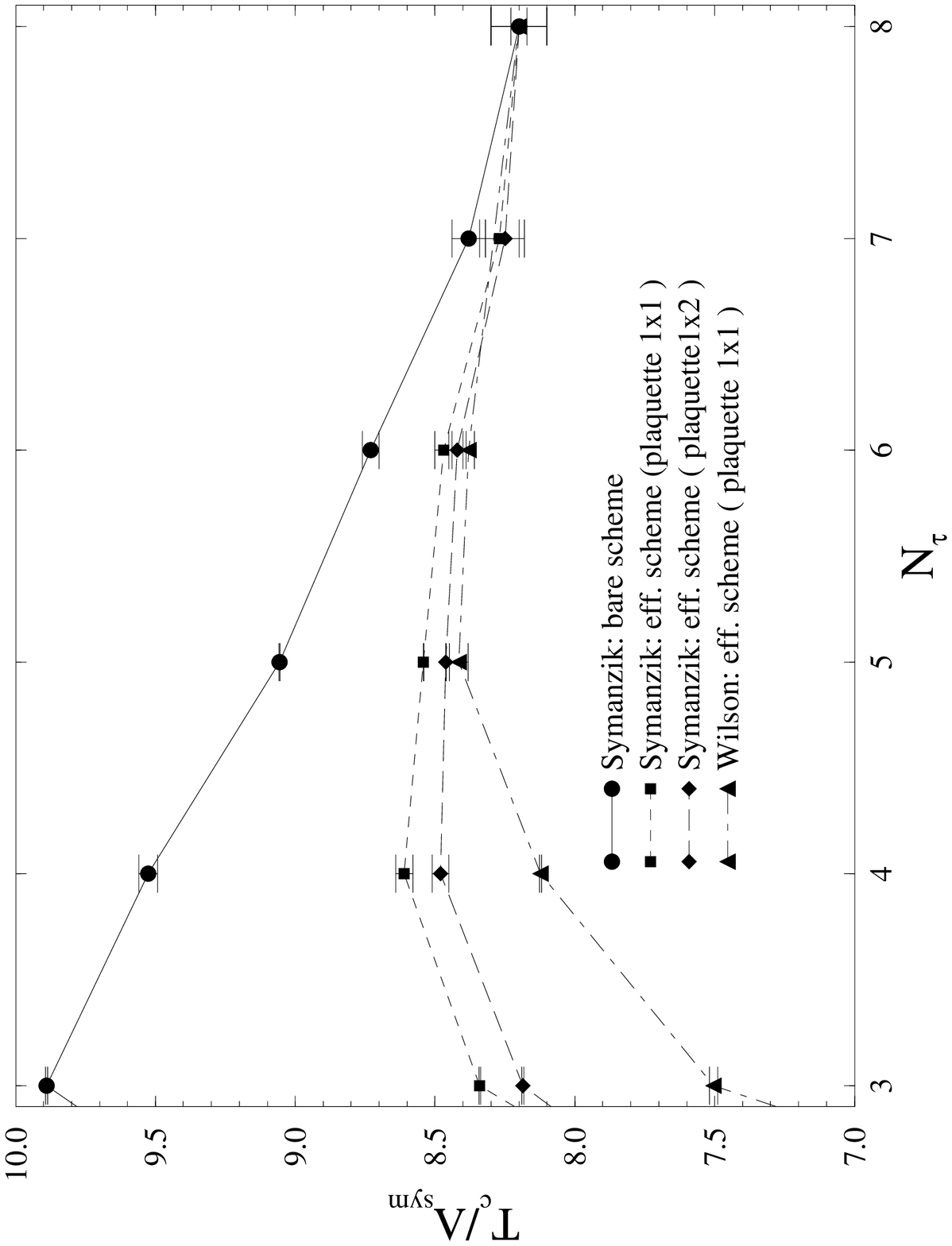}
}
\vskip 1.in
\caption{Comparison of ``bare'' and effective schemes}
\label{fig13}
\end{figure}
\clearpage

\begin{references}

\bibitem[*]{auth::cella} Electronic address: cella@sun10.difi.unipi.it
\bibitem[\dag]{auth::curci} Electronic address: curci@mvxpi1.difi.unipi.it
\bibitem[\ddag]{auth::lele} Electronic address: lele@vaxpia.pi.infn.it
\bibitem[\S]{auth::vicere} Electronic address: vicere@sun10.difi.unipi.it
\bibitem{fingberg:heller:karsch}
J.~Fingberg, U.~Heller and F.~Karsch, Nucl. Phys. B {\bf 392}, 493 (1993).

\bibitem{engels:fingberg:miller}
J.~Engels, J.~Fingberg and D.~E.~Miller, Nucl. Phys. B {\bf 387}, 501 (1992).

\bibitem{lepage:mackenzie}
G.~P.~Lepage and P.~B.~Mackenzie, preprint NSF-ITP-90-227 (1992).

\bibitem{parisi1}
G.~Parisi, Proceedings of the XXth
Conf. on High energy physics (Madison, WI 1980).

\bibitem{parisi2}
G.~Martinelli, G.~Parisi, and R.~Petronzio,
Phys. Lett. B {\bf 100}, 485 (1981).

\bibitem{curci:tripiccio}
G.~Curci and R.~Tripiccione, Phys. Lett. B {\bf 151}, 145 (1985).

\bibitem{symanzik:phi4}
K.~Symanzik, Nucl. Phys. B {\bf 226}, 187 (1983).

\bibitem{symanzik:sigma}
K.~Symanzik, Nucl. Phys. B {\bf 226}, 205 (1983).

\bibitem{parisi:sym}
G.~Parisi, Nucl. Phys. B {\bf 254}, 58 (1985).

\bibitem{symanzik:eff}
K.~Symanzik in New developments in gauge theories, ed. G. 't Hooft et al.
(Plenum, New York, 1980); in Lecture Notes in Physics {\bf 153}, ed. R.
Schrader et al. (Springer, Berlin, 1982); in Non-perturbative field theory
and QCD, ed. R. Jengo et al. (World Scientific, Singapore, 1983).

\bibitem{weisz}
P.~Weisz, Nucl. Phys. B {\bf 212}, 1 (1983).

\bibitem{curci:menotti:paffuti}
G.~Curci, P.~Menotti and G.~Paffuti, Phys. Lett. B {\bf 130}, 205 (1983).

\bibitem{luscher}
M.~L\"uscher and P.~Weisz, Commun. Math. Phys. {\bf 97}, 59 (1985),
erratum, Commun. Math. Phys.  {\bf 98}, 433 (1985).

\bibitem{engels:karsch:satz}
J.~Engels, F.~Karsch, H.~Satz and I.~Montvay, Nucl. Phys. B {\bf 205}[FS5],
545 (1982).

\bibitem{polyakov}
A.~M.~Polyakov, Phys. Lett. B {\bf 72}, 477 (1978).

\bibitem{svetitsky:yaffe}
B.~Svetitsky and G.~Yaffe, Nucl. Phys. B {\bf 210}[FS6], 423 (1982).

\bibitem{gutbrod:montvay}
F.~Gutbrod and I.~Montvay, Phys. Lett. B {\bf 136}, 411 (1984).

\bibitem{engels:fingberg:mitrjushkin}
J.~Engels, J.~Fingberg and V.~K.~Mitrjushkin, Phys. Lett. B {\bf 298}, 154
(1992).

\bibitem{barber}
M.~N.~Barber, in Phase Transition and Critical Phenomena Vol. 8,
ed.~C. Domb and J. Leibowitz, Academic Press (1981).

\bibitem{ape:machine}
A.~Bartoloni et al. ``An hardware implementation of the APE100
architecture'' Int. Journ. of Mod. Phys. C (1993) in press.

\bibitem{ape:language}
A.~Bartoloni et al. ``The software of the APE100 processor''
Int. Journ. of Mod. Phys. C (1993) in press.

\bibitem{kennedy:pendleton}
A.~D.~Kennedy and B.~J.~Pendleton, Phys. Lett. B {\bf 156}, 393 (1985).

\bibitem{ferrenberg:swendsen:1}
A.~M.~Ferrenberg and R.~H.~Swendsen, Phys. Rev. Lett.  {\bf 61}, 2635 (1988).

\bibitem{ferrenberg:swendsen:2}
A.~M.~Ferrenberg and R.~H.~Swendsen, Phys. Rev. Lett.  {\bf 63}, 1195 (1989).

\bibitem{weisz:wohlert}
P.~Weisz and R.~Wohlert, Nucl. Phys. B {\bf 236}, 397 (1984).
erratum, Nucl. Phys. B {\bf 247}, 544 (1984).

\bibitem{ma:wetzel}
J.~P.~Ma and W. Wetzel, Phys. Lett. B {\bf 176}, 441 (1986).

\bibitem{kovacs}
Eve Kovacs, Phys. Rev. D {\bf 25}, 871 (1982).

\bibitem{alves:berg:sanielevici}
N.~A.~Alves, B.~A.~Berg, S.~Sanielevici, Nucl. Phys. B {\bf 376}, 218 (1992).

\end{references}
\end{document}